\documentclass[12pt,preprint]{aastex}
\newcommand{\be}{\begin{equation}}
\newcommand{\ee}{\end{equation}}
\newcommand{\lb}[1]{\label{#1}}
\newcommand{\sty}{\scriptstyle}
\newcommand{\ssty}{\scriptscriptstyle}

\newcommand{\etal}{{\it et al.\ }}

\newcommand{\dl}{d_{\ssty L}}
\newcommand{\da}{d_{\ssty A}}
\newcommand{\dg}{d_{\ssty G}}
\newcommand{\dz}{d_z}

\newcommand{\an}{\langle n_0 \rangle}

\newcommand{\g}{\gamma}
\newcommand{\gest}{\gamma^\ast}
\shorttitle{Number Counts and Luminosity Function}
\shortauthors{Ribeiro and Stoeger}
\begin{document}
\title{Differential Density Statistics of Galaxy Distribution and
       the Luminosity Function}
\author{V.\ V.\ L.\ Albani,\altaffilmark{1} 
        A.\ S.\ Iribarrem,\altaffilmark{2} 
	M.\ B.\ Ribeiro~\altaffilmark{2} 
\affil{University of Brazil--UFRJ, Rio de Janeiro}
	and W.\ R.\ Stoeger~\altaffilmark{3}}
\affil{Vatican Observatory, University of Arizona, Tucson}
%\email{wstoeger@as.arizona.edu}
\altaffiltext{1}{ \ Mathematics Institute--UFRJ}
\altaffiltext{2}{ \ Physics Institute, University of Brazil - UFRJ, CxP
                 68532, CEP 21945-970, Rio de Janeiro, Brazil}
\altaffiltext{3}{ \ Vatican Observatory Group, Steward Observatory,
                 University of Arizona, Tucson, AZ 85721, USA}
\begin{abstract}
This paper uses data obtained from the galaxy luminosity function (LF)
to calculate two types of radial number densities statistics of the galaxy
distribution as discussed in Ribeiro (2005), namely the differential
density $\gamma$ and the integral differential density $\gamma^\ast$.
By applying the theory advanced by Ribeiro and Stoeger (2003), which
connects the relativistic cosmology number counts with the astronomically
derived LF, the differential number counts $dN/dz$ are extracted from
the LF and used to calculate both $\gamma$ and $\gamma^\ast$ with
various cosmological distance definitions, namely the area distance,
luminosity distance, galaxy area distance and redshift distance. LF
data are taken from the CNOC2 galaxy redshift survey and $\gamma$ and
$\gamma^\ast$ are calculated for two cosmological models: Einstein-de
Sitter and an $\Omega_{m_0}=0.3$, $\Omega_{\Lambda_0}=0.7$ standard
cosmology. The results confirm the strong dependency of both statistics
on the distance definition, as predicted in Ribeiro (2005), as well as
showing that plots of $\gamma$ and $\gamma^\ast$ against the luminosity
and redshift distances indicate that the CNOC2 galaxy distribution follows
a power law pattern for redshifts higher than 0.1. These findings bring
support to Ribeiro's (2005) theoretical proposition that using
different cosmological distance measures in statistical analyses of
galaxy surveys can lead to significant ambiguity in drawing conclusions
about the behavior of the observed large scale distribution of galaxies. 
\end{abstract}

%\keywords{cosmology: number counts, luminosity function, relativity}

\section{Introduction}\lb{intro}

The {\it galaxy luminosity function} (LF) is an important analytical
tool for testing cosmological models. It conveys observational
information on how the local density of galaxies varies for
different types of galaxies and with differing cosmic environments,
as well as how the local galaxy density changes with cosmic epoch.
The LF has been extensively and systematically calculated for a great
variety of galaxy redshift surveys, in terms of the number of galaxies,
their morphological types, observational bandwidths and redshift
depths. In addition, since in calculating the LF one often includes a
model of source evolution, the presently available LF parameters
obtained from different galaxy samples provide an important collection
of data for testing cosmological models. To do so, however, we need a
detailed theoretical framework connecting the astronomical LF data
calculation and practice to cosmological modeling. This is essential
for extracting information that can be used for further constraining
cosmological models. 

In a recent paper, Ribeiro and Stoeger (2003; from now on RS03) have
presented a detailed relativistic treatment directly linking the
cosmology source counting theory with LF data. 
RS03 started with a general relativistic treatment of number source
counts in a general spacetime and then derived expressions for relativistic
density per source and total relativistic energy densities of the
universe in terms of the luminosity and selection functions. These
theoretical/observational relationships were applied to test the
LF parameters determined from the CNOC2 galaxy redshift survey (Lin
\etal 1999) for consistency with the assumed Einstein-de Sitter (EdS)
cosmological model. Other possible applications and extensions of this
general framework were also outlined.

This present work explores one of these extensions of RS03. First, we
shall show how to extract the differential number counts $dN/dz$ from
LF data. Inasmuch as calculating the LF parameters from a galaxy sample
requires the assumption of a cosmological model, since a volume
definition, usually comoving, must be employed, one can factor the
cosmology out of the LF parameters and re-obtain the ``raw'' number count
data. The methodology for determining LF's requires fitting the original
data to some analytical form of the LF, usually of the Schechter type,
where a model of source evolution is also included. Therefore, by
```raw' number count data'' we mean that the reverse process eliminates
the cosmology, but does not remove the fitting itself, that is, the LF
analytical assumption and source evolution corrections.

Once these ``raw'' $dN/dz$ are in our possession, we carry out a
statistical analysis with the differential density parameters 
discussed in Ribeiro (2005), which are basically different forms of
number densities. As such this analysis is based on the differential
number counts. Thus, it can be done in any cosmological model (since
the $dN/dz$ data does not have any kind of cosmology built into it).
Here we test the theoretical proposition advanced in Ribeiro (2005) that
using different cosmological distance measures in statistical analyses
of galaxy surveys can lead to significant ambiguity in drawing
conclusions about the behavior of the observed large scale distribution
of galaxies. 

Besides the sensitivity of these large scale analyses to the cosmological
distances used, comparison of number counts at different redshifts is made
more difficult by the fact that, as we look down our past light cone, we are
counting galaxies at different redshifts, and therefore at different times.
Our line of sight cuts across many different space-like, possibly nearly
spatially homogeneous, surfaces. This means, that, even aside from the
distance measure employed, the number counts down our light cone will
not correspond in any simple way with the integrated galaxy number density
over a volume on any space-like three-surface.

Thus, even though there is ``almost homogeneity'' \footnote{ \ We use
this term to describe succinctly one of the key characteristics of a
perturbed Friedmann-Lema\^{i}tre-Robertson-Walker (FLRW) universe. Our
universe is obviously not exactly spatially homogeneous -- otherwise we
would not be here! But there is strong evidence that it is ``almost
homogeneous,'' or ``almost FLRW,'' on very large scales.} of the
distribution of galaxies, and therefore of mass-energy, on spatial
3-surfaces (Peebles 1993 and references therein, Hogg et al.\ 2005), we
should not expect to find ``observational homogeneity'' as we count
galaxies down our past light cone. Yet, employing some distance measures
(e.g. the co-moving distance and the galaxy area distance) leads to the
appearance of such observational homogeneity. For other measures of
distance, we obtain, as Ribeiro (2005) has shown theoretically and as
we here shall show from the CNOC2 data, a power-law galaxy distribution
down the light cone, indicating a {\it possible} self-similar, hierarchical
clustering or fractal-like pattern -- even though the actual spatial
distribution is almost homogeneous. This is simply the resulting effect
of collecting data from successively earlier epochs as we go to larger
redshifts.

Besides showing how to extract cosmology-free $dN/dz$ information
from LF's, a primary objective of the paper is to clarify these issues,
emphasizing the illusory and ambiguous character of such observational
homogeneity and of such purely observational galaxy distributions, in
general.
 
The main conclusions we reach in this paper are as follows.
%(1)
The galaxy distribution given in the CNOC2 redshift
survey supports Ribeiro's (2005) theoretical finding that
the statistics of the observed galaxy distribution are strongly
dependent on the choice of cosmological distance definitions.
%(2)
Both the differential density $\gamma$, which gives the rate of
increase in galaxy number counts with distance down the past light
cone, and the integral differential density $\gamma^\ast$, which
is simply the integral of $\gamma$ over a specified volume, decay
with higher luminosity and redshift distances, with no flattening
of this decrease being found within the limits of the CNOC2 survey
($0 \le z \le 1$).
%(3)
These decreasing values of the differential densities show a
tendency to linearization in a log-log plot for $z>0.1$, that is, a
power law pattern, which suggests a self-similar behavior for
the observed galaxy distribution in those ranges. Again, it is
important to recognize that these parameters are measuring changes
in galaxy number density down our past light cone, {\it not} in
the space-like surfaces of the Friedmann-Lema\^{i}tre-Robertson-Walker
(FLRW) models where there is evidence of ``almost homogeneity''. These
results are valid for both EdS cosmology and the currently favored FLRW
standard model with $\Omega_{m_0}=0.3$ and $\Omega_{\Lambda_0}=0.7$.
However, by comparing the results of both models it seems that
%(4)
the linearization tendency mentioned above is more pronounced
in the second model, suggesting that FLRW cosmologies
with a less than critical matter density parameter might be a better
fit as far as a self-similar distribution is concerned.

The plan of the paper is as follows. In \S 2 the main equations,
derived in RS03, necessary for our purposes here are summarized and
the method for extracting the differential number count $dN/dz$
from LF data is developed and applied to the CNOC2 galaxy redshift
survey data. In \S 3 we calculate the differential density statistics
in two cosmological models, namely EdS and the presently favored
standard model with $\Omega_{m_0}=0.3$, $\Omega_{\Lambda_0}=0.7$.
\S 4 summarizes our conclusions and the appendix briefly
describes the error analysis performed with the CNOC2 data.

\section{Differential Number Counts from LF Data}\lb{lfdata}

In this section we present a summary of the relevant
results from RS03 and derive the general equation for the
observed differential number count. Then we will specialize these
equations to the EdS model in order to be able to obtain the ``raw''
$dN/dz$ from the LF parameters calculated with the CNOC2 galaxy
redshift survey.

\subsection{General Relations}\lb{gen}

We start with the general equation for the {\it number of
cosmological sources} $N$ in a volume section at a certain point down
the null cone whose {\it affine parameter distance} is given by $y$,
\be
 dN={\left( \da \right) }^2 \left( 1+z \right) d \Omega_0 \; n(y) dy.
 \lb{um}
\ee
Here $\da$ is the {\it area distance} (also known as {\it
angular diameter distance}, {\it observer area distance} or {\it
corrected luminosity distance}) of this section from the observer's
viewpoint, $d \Omega_0$ is the {\it solid angle} of the volume
section at the observer's position, $n$ is the {\it number density} of
radiating sources per unit proper volume and $z$ is the {\it redshift}.
This equation is simply a more convenient rewriting of Ellis' (1971)
key result (see also Pleba\'{n}ski and Krasi\'{n}ski 2006) for
cosmological number counts in any cosmological model (see details in
RS03, \S 2.1.1). The number density $n$ appearing in the equation
above is related to the {\it galaxy luminosity function} $\phi$ as
follows,
\be n(y)=\Psi[z(y)]=\int_0^\infty \phi(l)dl,
    \lb{dois}
\ee
where $l=L/L_\ast$, $L$ is the absolute luminosity of the source and
$L_\ast$ is the luminosity scale parameter (RS03, eqs.\ 29, 35).
In addition we have that
\be \rho = \mathcal{M}_g \Psi,
    \lb{tres}
\ee
where $\rho$ is the {\it local density} as given by the right-hand side
of Einstein's field equations and $\mathcal{M}_g$ is the {\it average
galaxy rest mass} (RS03, \S 2.1.2, eq.\ 40). Therefore, equation 
(\ref{um}) can be written as,
\be
  \frac{dN}{dz}= { \left( \da \right) }^2 \left( 1+z \right) d \Omega_0
                 \left( \frac{\rho}{\mathcal{M}_g} \right)
		 \frac{dy}{dz}.
  \lb{quatro}
\ee

This equation can be directly written in terms of observable
quantities in a certain bandwidth $W$ as follows (RS03, \S 4.4,
eqs.\ 46, 84),
\be { \left[ \frac{dN}{dz} \right] }_0 = \sum_W a_{\ssty W} { \left[
     \frac{dN}{dz} \right] }^{\ssty W}.
     \lb{cinco}
\ee
Here the subscript zero means an observed quantity, and the superscript
index ${\sty W}$ means the observational bandwidth used when the data
was collected. The factor $a_{\ssty W}$ is the fraction of galaxies in a
certain wave band that are not counted in other bandwidths. This is
necessary to avoid overcounting due to possible overlap of galaxy
counts in different bandwidths. Therefore, we have that (RS03, eqs.\
45, 46),
\be a_{\ssty W}(z) =
    \left\{
    \begin{array}{ll}
    1, & \mbox{for} \; \; \;  W=1, \\
    b_{\ssty W}(z) < 1, & \mbox{for} \; \; \; W>1.
    \end{array}
    \right.
    \lb{aw}
\ee
where, $b_{\ssty W}(z)$ is the fraction of galaxies in each waveband,
that is, when $W>1$, which are not counted in wavebands $1,2,\ldots,(W-1)$.
Thus, equation (\ref{quatro}) can also be written 
\be { \left[ \frac{dN}{dz} \right] }_0 =
       { \left( \da \right) }^2 \left( 1+z \right) d \Omega_0 \;
       \frac{dy}{dz}  { \left[
       \frac{{\rho}}{{\mathcal{M}_g}} \right] }_0= 
       { \left( \da \right) }^2 \left( 1+z \right) d \Omega_0 \;
       \frac{dy}{dz}  \sum_{\ssty W} a_{\ssty W}
       { \left[ \frac{{\rho}}{{\mathcal{M}_g}} \right]
       }^{\ssty W}. 
    \lb{dndz0}
\ee
Similarly, equation (\ref{tres}) yields
\be \rho_0={ \left[ \Psi \mathcal{M}_g \right] }_0
          =\sum_{\ssty W} a_{\ssty W} { \left( \psi^{\ssty W}
	   \mathcal{M}_g \right) }
	  =\sum_{\ssty W} a_{\ssty W} \sum_v P_v^{\ssty W} \mathcal{M}_v
	  \psi_v^{\ssty W}.
    \lb{rho0}
\ee
Here $\psi^{\ssty W}$ is the {\it selection function} calculated from
the luminosity function obtained in a certain wave band $W$,
\be
  \psi^{\ssty W}[l(z)]=\int_{l(z)}^\infty \phi^{\ssty W} (l) \; dl,
  \lb{psi}
\ee
$P_v^{\ssty W}$ is the {\it galactic morphology population fraction},
giving the abundance of each galactic type relative to the total number
of counted galaxies in each redshift range and observed in a certain
wave band $W$. Finally, $\mathcal{M}_v$ is the average rest-mass 
for galaxies of a certain morphological class $v$ (see eqs.\ 13, 14, 87
of RS03).

{From} the discussion above it is clear that it is still possible to
write the observed ratio between the local density and the average
galaxy rest-mass as follows,
\be { \left[ \frac{{\rho}}{{\mathcal{M}_g}} \right] }^{\ssty W} =
    \frac{ \sum_v P_v^{\ssty W} \mathcal{M}_v \psi_v^{\ssty W}}{
    \sum_v P_v^{\ssty W} \mathcal{M}_v}.
    \lb{rmgw}
\ee
This equation allows us to write the observed differential number
counts in terms of the chosen cosmological model and observational
quantities extractable from galaxy catalogues. This result may be
written as below,
\be { \left[ \frac{dN}{dz} \right] }_0 =
       { \left( \da \right) }^2 \left( 1+z \right) d \Omega_0 \;
       \frac{dy}{dz}  \sum_{\ssty W} a_{\ssty W}
       \left[
    \frac{ \sum_v P_v^{\ssty W} \mathcal{M}_v \psi_v^{\ssty W}}{
    \sum_v P_v^{\ssty W} \mathcal{M}_v}
       \right].
    \lb{dndz02}
\ee
Notice again that this equation is absolutely general, independent of
cosmological model and independent of the survey, since its
observational part under the summation signs is written in such a
way as to be applicable, in principle, to any galaxy sample
(RS03, \S 3.1).

\subsection{Evolution of Galaxy Masses}

A straightforward examination of equation (\ref{dndz02}) shows that
it depends on the typical average rest-mass $\mathcal{M}_v$ of
galaxies belonging to the morphological class $v$. The way $\mathcal{M}_v$
appears in this equation might indicate that this typical mass remains
unchanged for different redshifts, a hypothesis which is certainly
unrealistic as there is today evidence for galaxy mergers. Therefore, we
must consider some dependency of the typical galaxy mass with the
redshift in order to have a realistic assumption.

One way of doing this would be by replacing $\mathcal{M}_v$ in equation
(\ref{dndz02}) by $[ f(z) \: \widetilde{\mathcal{M}}_v ]$, where
$\widetilde{\mathcal{M}}_v$ is the typical mass of $v$ type galaxy at
$z=0$, that is, at rest with us, or here and now, and such that the function
$f(z)$ obeys $f(0)=1$. So, if a galaxy doubles its mass from $z=1$ to $z=0$
we could take $f(z)=(1+z)^{-1}$. A milder growth would be achieved by
assuming $f(z)=(1+z)^{-q}$, with $0<q<1$. So, for $q=0.5$ the galaxy would
have 70\% of its present mass when $z=1$, whereas with $q=0.2$ we go from
$0.87 \: \widetilde{\mathcal{M}}_v$ at $z=1$ to the present ($z=0$)
rest-mass value $\widetilde{\mathcal{M}}_v$.

There is support for this kind of galaxy mass evolution as Carlberg
\etal (2000) showed that in the CNOC2 and in the Caltech Faint Galaxy
Survey galaxies of mass within an order of magnitude of
${\mathcal{M}_\ast}$ will grow by about $0.15 \: {\mathcal{M}_\ast}$
from $z=1$ to $z=0$. Here ${\mathcal{M}_\ast}$ is the typical mass of a
galaxy with luminosity given by the luminosity scale parameter $L_\ast$
of Schechter's LF (eq.\ \ref{dois}). Simulations carried out by Murali
\etal (2001) agree with Carlberg \etal (2000). Both studies emphasize,
however, that these results only pertain to galaxies which are large,
that is, with masses near ${\mathcal{M}_*}$, leaving out low mass
galaxies which should also grow by mergers and accretion, but not as
much as the larger galaxies. Both studies do not mention possible
dependency of galaxy mass evolution with morphological types. Therefore,
a mild growth such as setting $q=0.2$ for all galaxy morphologies
seems a quite reasonable first order approximation considering all
these uncertainties.

At this point one must look again at equation (\ref{dndz02}) to notice
that if we replace  $\mathcal{M}_v$ by $[ f(z) \:
\widetilde{\mathcal{M}}_v ]$ the mass evolution estimate $f(z)$ is
canceled out in that equation. This means that we can use the values
of the typical masses at rest with us in equation (\ref{dndz02}),
since, in that expression, the following approximation holds:
$\mathcal{M}_v \simeq \widetilde{\mathcal{M}}_v$. In other words,
the observed differential number counts (\ref{dndz02}) constructed
from LF parameters, which is the basic quantity to be used in all
analyses of this paper, is, therefore, not affected by galaxy mass
evolution, at least to first-order approximation.

Notice that although galaxy mass evolution does not directly change
equation (\ref{dndz02}), it appears indirectly since the selection
functions in equation (\ref{dndz02}) were obtained from LF parameters
whose derivation usually include some sort of source evolution.

\subsection{Einstein-de Sitter Cosmology}

We now specialize the equations above to Einstein-de Sitter (EdS) cosmology.
This is done by finding the equations for the affine parameter and the area 
distance in terms of the redshift. It is simple to show that both
equations can be written (RS03, \S 4.2) 
\be \frac{dy}{dz}=\frac{c}{H_0 {\left( 1+z \right) }^{7/2}},
    \lb{dydz}
\ee
\be
  \da = \frac{2c}{H_0} \frac{\left( 1+z-\sqrt{1+z} \right)}{{\left(
        1+z \right) }^2},
  \lb{da}
\ee
where $H_0$ is the Hubble constant and $c$ is the light speed. 
Throughout this paper we take $H_0=100 \; \mathrm{km} \; \mathrm{s}^{-1}\;
\mathrm{Mpc}^{-1}$ as this is the value adopted by Lin \etal (1999).
We aim here to factor the cosmology out of LF data and since the
value of the Hubble constant is embedded in LF calculations, we must
adopt the same value originally set for the LF calculation. In the
case of this paper it then must be the same as used by Lin \etal (1999).

An important issue in LF calculations is the volume definition.
Currently observers adopt the comoving volume, whereas all equations
above assume the proper volume. So, we need a volume conversion,
obviously dependent on the cosmological model, in order to actually
use the selection function in our equations above. Thus, in the EdS model
the selection functions in both these volumes are related by 
(RS03, footnote 6 and \S 4.3), 
\be
   \psi_v^{{\ssty W}
   \! \! \! \! \! \! \! \! \! \! \! \! \! \! \! \! \! \! \!
   \mathrm{\ssty PR}}  =
   \frac{9}{4} { \left( H_0 \right) }^2 {\left( 1+z \right) }^3 \; \; \;
   \psi_v^{{\ssty W}
   \! \! \! \! \! \! \! \! \! \! \! \! \! \! \!
   \mathrm{\ssty C}},
   \lb{psis}
\ee
where the left superscripts PR and C denote that the selection function
is evaluated with the proper and comoving volumes, respectively.

Bearing in mind the results above and spherical symmetry, we are now
in position to write the observed differential number counts
(\ref{dndz02}) in terms of the selection function in the EdS
cosmological model as follows,
\be { \left[ \frac{dN}{dz} \right] }_0 = \left(
    \frac{36 \pi c^3}{H_0} \right) \frac{ {\left( 1+z-\sqrt{1+z}
    \right)}^2}{{\left( 1+z \right)}^{7/2}} \sum_{\ssty W}
    a_{\ssty W} \left[ \frac{ \sum_v P_v^{\ssty W} \mathcal{M}_v
       \; \; \; \psi_v^{{\ssty W} \! \! \! \! \! \! \! \!
       \! \! \! \! \! \! \! \mathrm{\ssty C}}}
       { \sum_v P_v^{\ssty W} \mathcal{M}_v}
       \; \right].
    \lb{dndzeds}
\ee

\subsection{$\mathbf{dN/dz}$ from LF Data of CNOC2 Galaxy Survey}\lb{cnoc2}

The CNOC2 sample was obtained in three wave band filters: $R_c$, $U$,
$B_{\ssty AB}$ (hereafter, we shall for notational ease, abbreviate
the latter as ``B''). This gives us the possible values for the wave
band index $W=R_c$, $U$, $B$. So, according to equation (\ref{aw}) we
have $a_{\ssty W}= b_{\ssty W}$ and since all galaxies appear in all
bandwidths, we have that $$b_{\ssty R_c}=b_{\ssty B}=b_{\ssty U}=1/3.$$
This catalogue has morphological indexes $v=1,2,3$ denoting the three
morphological types in which the sample population was divided, early,
intermediate and late spectral types, respectively (Lin \etal 1999).
These galaxy morphological types have very approximate dynamical
average rest masses equal to
$\mathcal{M}_1=0.5 \times 10^{11} \mathcal{M}_\odot$,
$\mathcal{M}_2=0.3 \times 10^{11} \mathcal{M}_\odot$ and
$\mathcal{M}_3=0.1 \times 10^{11} \mathcal{M}_\odot$ (Sparke and
Gallagher 2000, pp.\ 204 and 264). As discussed in RS03, the CNOC2
sample is such that the proportion $P_v$ of each galaxy type population
is as follows (RS03, \S 3.4, tables 1-3), 
$$ \left\{
    \begin{array}{llll}
   P_1^{\ssty R_c}  = & P_1^{\ssty U}  = & P_1^{\ssty B}  = & 0.29, \\ 
   P_2^{\ssty R_c}  = & P_2^{\ssty U}  = & P_2^{\ssty B}  = & 0.24, \\ 
   P_3^{\ssty R_c}  = & P_3^{\ssty U}  = & P_3^{\ssty B}  = & 0.47. 
    \end{array}
    \right. 
    \lb{pv}
$$
With these results equation (\ref{dndzeds}) can be applied to the CNOC2
sample, yielding,
\be { \left[ \frac{dN}{dz} \right] }_0 = \left(
    \frac{36 \pi c^3}{{ H_0 }^3} \right) \frac{ {\left( 1+z-\sqrt{1+z}
    \right)}^2}{{\left( 1+z \right)}^{7/2}} \mathop{\sum_{\ssty
    W=1}^3}_{ \left( \ssty R_c, U, B \right) }b_{\ssty W} \left[
    \frac{ \sum_{v=1}^3 P_v^{\ssty W} \mathcal{M}_v \left(
       \; \; \; \; \; \; \; \: \psi_{v \! \! \! \! \! \! \! \! \! \! 
       \! \! \! \! \! \! \! \! \! \! \! \! \! \!\! \! \mathrm{\ssty 
       CNOC2}}^{{\ssty W}\! \! \! \! \! \! \! \! \! \! \! \! \! \! \! 
       \mathrm{\ssty C}} \; \; \: \right) }{ \sum_{v=1}^3 P_v^{\ssty W}
       \mathcal{M}_v}\right],
    \lb{dndzcnoc2}
\ee
where the correction 
$$ \left( \; \; \; \; \; \; \; \; \psi_{v \! \! \! \! \! \! \!
   \! \! \! \! \! \! \! \! \! \! \! \! \! \! \! \! \!\! \! \mathrm{\ssty 
   CNOC2}}^{{\ssty W}\! \! \! \! \! \! \! \! \! \! \! \! \! \! \! 
   \mathrm{\ssty C}} \; \; \: \right) = {\left( H_0 \right) }^2 \; 
   \left( \; \: \psi_v^{{\ssty W} \! \! \! \! \! \! \! \! \! \! \! \! \! \! \!
   \mathrm{\ssty C}} \; \right)
$$
is necessary in order to have correct units. Considering now
the CNOC2 figures above, this equation can still be changed to,
\begin{eqnarray}
    { \left[ \frac{dN}{dz} \right] }_0 & = & \left(
    \frac{36 \pi c^3}{{H_0}^3} \right) \frac{ {\left( 1+z-\sqrt{1+z}
    \right)}^2}{{\left( 1+z \right)}^{7/2}}
    \frac{1}{7.92 \times 10^{10}} \times \nonumber \\ 
    & & \times \Big[
    \left(
    \psi_1^{\ssty R_c} + \psi_1^{\ssty U} + \psi_1^{\ssty B}
    \right) 1.45 \times 10^{10} + \nonumber \\ 
    & & 
    + \; \: \left( 
    \psi_2^{\ssty R_c} + \psi_2^{\ssty U} + \psi_2^{\ssty B}
    \right) 7.2 \times 10^{9} + \nonumber \\ 
    & &
    + \; \: \left(
    \psi_3^{\ssty R_c} + \psi_3^{\ssty U} + \psi_3^{\ssty B}
    \right) 4.7 \times 10^{9} 
    \Big], 
    \lb{dndzpsi}
\end{eqnarray}
where the C and CNOC2 indexes of the selection function in equation
(\ref{dndzcnoc2}) were dropped. The observed differential number count
of the CNOC2 redshift survey can then be finally calculated. Tables
\ref{table1}, \ref{table2} and \ref{table3} below reproduce the
numbers calculated and presented in RS03 with errors calculated here
(see \S \ref{app}), which are enough to evaluate equation
(\ref{dndzpsi}). The results are shown in table \ref{table4}.

Notice that the value of the Hubble constant in equation (\ref{dndzpsi})
must be the same as originally used by Lin \etal (1999) in deriving the
CNOC2 LF, that is, $H_0=100 \; \mathrm{km} \; \mathrm{s}^{-1}\;
\mathrm{Mpc}^{-1}$.

\section{Differential Density Statistics with CNOC2 Derived Data}\lb{gamas}

We shall now present the differential density radial statistics required
to analyze the galaxy number density distribution and its conceptual
implications. Some of this was previously discussed in Ribeiro (2001,
2005) and, therefore, \S \ref{gr2} is in part a review. Subsequently we
will apply these general expressions to the CNOC2 data obtained in the
previous section in two different standard cosmological models.

\subsection{Basic Equations and Concepts}\lb{gr2}

\subsubsection{Definitions}

The {\it differential density} $\gamma$ at a certain {\it observed
distance} $d_0$ is defined by the following expression,
\be \gamma = \frac{1}{S_0} \frac{dN}{d (d_0) },
    \lb{gamma}
\ee
where
\be S_0 = 4 \pi { \left( d_0 \right) }^2
    \lb{s0}
 \ee
is the {\it area of the observed shell} of radius $d_0$.
This definition appears to have been initially used in the context
of the galaxy distribution by Wertz (1970, 1971). As can be easily
seen, $\gamma$ is really very much like the similar differential
number counts parameter $dN/dz$ (see eq.\ 4), except that it
is differential with respect to an observable distance down our
past light cone, instead of with respect to $z$ (which is also
an observable parameter down our past light cone). Furthermore, it
is ``normalized'' with respect to the ``observed surface area'' $S_0$.
Essentially, then, $\gamma$ tells us the rate of growth in the
number counts, or more exactly in their density, as one moves
down the past light cone in the observable distance $d_0$. As
is obvious, the behavior of this parameter will depend heavily
on which distance we employ for $d_0$ -- observer area distance, 
galaxy area distance, luminosity distance, redshift distance, or
something else. 

Furthermore, Ribeiro (2005) has advanced an extension of $\gamma$
called the {\it integral differential density} $\gamma^\ast$,
which is simply the integration of $\gamma$ over the {\it observed
volume} $V_0$. Such a definition yields, 
\be
   \gamma^\ast= \frac{1}{V_0} \int_{V_0} \gamma \; dV_0,
   \lb{gammas}
\ee
where
\be V_0=\frac{4}{3} \pi { \left( d_0 \right) }^3. \lb{v0} \ee
We should carefully note that $dV_0$ is different from the volume
element given in the usual form of the metric. This is because
it involves $d(d_0)$ down our past light cone, instead of something
like $a(t)dr$ at one given time $t = t_1$ -- that is on space-like
surface, usually one of ``almost homogeneity.''

It is more useful to write $\gamma$ and $\gamma^\ast$ in terms of the
redshift, as follows,\footnote{ \ This notation for the differential
densities is slightly different than in Ribeiro (2005), since here
we follow RS03 very closely and adopt the subscript index zero to
refer to observed quantities, built from galaxy sample derived
observations as much as possible, whereas the unindexed quantities
refer to theoretical expressions, built from theory.}
\be \gamma(z) = \frac{dN}{dz} { \left[ S_0(z) \; \frac{d}{dz}
                \left[ d_0 (z) \right] \right] }^{-1},
    \lb{gamma2}
\ee
\be \gamma^\ast (z)=\frac{1}{V_0(z)} \int_0^z \gamma (z) \;
    \frac{dV_0}{dz} dz.
    \lb{gammas2}
\ee
The equations above can be obtained from any given cosmological model,
since all quantities on the right hand side can be calculated from
the geometry of the model and the assumed matter distribution. The
equivalent equations built from data derived galaxy samples should
then read as follows,
\be \gamma_0 (z) = { \left[ \frac{dN}{dz} \right] }_0 { \left[ S_0 \;
    \frac{d}{dz}
                (d_0) \right] }^{-1},
    \lb{gamma3}
\ee
\be \gamma_0^\ast (z) =\frac{1}{V_0} \int_0^z \gamma_0  \;
    \frac{dV_0}{dz} \; dz.
    \lb{gammas3}
\ee
Notice that by their definitions in equations (\ref{gamma}),
(\ref{s0}), (\ref{gammas}), (\ref{v0}), the following result holds,
\be
  \gamma^\ast = \frac{N}{V_0} = \langle n \rangle,
  \lb{nmed}
\ee
where $\langle n \rangle$ is the average number density.

\subsubsection{Conceptual Implications}

\subsubsubsection{Spatial and Observational Homogeneities}\lb{soh}

In order to interpret the meaning and significance of $\gamma$ and
$\gamma^\ast$ further, we emphasize that in cosmological models we 
can define {\it two} distinct types of homogeneity.
{\it Spatial homogeneity} is defined in the usual sense, that is,
the usual spatial density $\rho$ appearing on the right hand side of
Einstein's field equations is dependent only on time, $\rho=\rho(t)$,
which means that in  each constant time hypersurface the local density
remains unchanged. So, {\it spatial homogeneity means that quantities
such as density are constant on a space-like foliation of surfaces.} The
primary example, of course, are the surfaces of homogeneity in FLRW.

However, there is another type of homogeneity, {\it observational
homogeneity}, that can also be defined in cosmology. It is the property
that some average density calculated with observational data derived
from galaxy redshift surveys is constant. This is sometimes what one
has in mind when one talks about the possible homogeneity of the galaxy
distribution in the context of observational cosmology. So,
{\it observational homogeneity is defined in terms of average
observational quantities along our past like cone, instead of in a
given space-like surface}. Operationally, the {\it observational}
average density ${\langle \rho_0 \rangle}$ is constant.\footnote { \ The
notation may be a bit confusing here, as $\rho_0$ is the local density
at the present constant time $t_0$, whereas ${\langle \rho_0 \rangle}$
is the average of $\rho[t(r)]$ along the null cone $t(r)$, since
$\rho[t(r)]$ will change along the past null cone (see Ribeiro 2001
for in depth discussions of this topic).} How to relate ${\langle
\rho_0 \rangle}$ to the above-mentioned $\rho(t)$, how to determine
$\rho (t)$ from an observed ${\langle \rho_0 \rangle}$, and what the
relationship between observational homogeneity and spatial
homogeneity are key issues.

Since a basic method for observationally dealing with galaxies is
by counting them, it is reasonable to state formally that {\it
observational homogeneity is achieved when the average number density
$\an$ remains constant, as one averages over different volumes 
down our past light cone}. So, 
\begin{equation}
  \an \equiv \frac{N_0}{V_0}= \mbox{constant},
  \label{exchange}
\end{equation}
where $N_0$ is the number of observed galaxies counted within a
certain volume $V_0$. As indicated above, this definition of
observational density implies that both $N_0$ and $V_0$ are obtained
along our past light cone. So, equation (\ref{exchange}) {\it is not
applicable over space-like surfaces.}

The practical calculation of $\an$ by means of equation (\ref{exchange})
encounters some basic operational difficulties. We immediately 
face the problem of identifying the volume $V_0$. Actually, since $N_0$ is
observed, $V_0$ should also be an observed quantity. But, we cannot yet
translate the observed redshift $z$ into a cosmological distance without
a definite cosmological model. So, we must choose a certain distance
measure among several possibilities from the model (many workers often
choose the comoving distance), observe $N_0$ and calculate $\an$.
Therefore, although spatial homogeneity can be uniquely defined relative
to a space-like slicing, observational homogeneity cannot be. Obviously,
{\it a model can satisfy spatial homogeneity, without satisfying 
observational homogeneity,} or vice-versa. In fact this will often be
the case.

This was shown to be correct in Ribeiro (2001, 2005), with earlier 
discussions in  Ribeiro (1992, 1993, 1995). Since we can also obtain
$N$ from theory, Ribeiro (2005) showed that, although EdS cosmology
is spatially homogeneous by definition, it may or may {\it not} be
observationally homogeneous, depending on the particular type of
distance chosen. Thus, according to Ribeiro (2005), an
FLRW model can satisfy {\it both} types of homogeneity, spatial
{\it and} observational, {\it if, and only if}, one adopts the
galaxy area distance, or comoving distance, in the definition of
observational homogeneity, as given by equation (\ref{exchange}).
Furthermore, it is clear that lack of observational homogeneity is
perfectly consistent with spatial homogeneity, and that there will
be cases of observational homogeneity in which there is no spatial
homogeneity (Rangel Lemos and Ribeiro 2006, in preparation). In
specifying the link between the two the distance being used is
crucial!

From this point of view, interpreting $\g$ and $\gest$ is straightforward.
Considering equations (\ref{nmed}) and (\ref{exchange}) it is clear
that $\gamma^\ast_0 = \an$. This means that $\g$ is a useful tool
for determining $\an$. But this is true only if we are trying to 
determine whether or not the galaxy number count $N$ is
{\it observationally} homogeneous.

\subsubsubsection{Other Remarks}

It is important to point out that only one of the two factors on the
right hand side of equation (\ref{gamma3}), can, at present, be
calculated directly from observations. Only the observed
differential count ${ \left[ dN/dz \right] }_0$ can be evaluated
from equation (\ref{dndz02}) or one of its specialized forms
(\ref{dndzeds}), (\ref{dndzcnoc2}), (\ref{dndzpsi}).\footnote{ \ The
observed differential number count ${ \left[ dN/dz \right] }_0$ does
not need to be estimated only from LF data, as it can be directly
observed. Nevertheless, in the context of this paper, we are
recovering it from the LF's, since those are often what is made
available by the observers.}\label{note6} To calculate the second
factor of equation (\ref{gamma3}), the {\it geometrical} part involving
$S_0$ and $d_0$, one must still assume a cosmological model, inasmuch
as obtaining observational distances, either by standard candles or
standard rods, of every galaxy in a redshift survey is currently
beyond our observational capabilities.

Along with this, as we have already indicated above and in the
introduction, there is the key problem that the observational
distance $d_0$ as obtained from theory is {\it not} unique, meaning
that the observed density of the galaxy distribution is dependent on
this distance choice. This applies even both to $\gamma$ and to
$\gamma^\ast$ -- $\gamma^\ast$ depends on the volume $V_0$ over which
the galaxies are counted, which in turn depends on the distance choice.
We have already seen that that this choice can make a significant
difference. Which distance is appropriate?

As it has been extensively argued elsewhere, these are the basic,
but, so far, not generally acknowledged, dilemmas underlying the
observational determination of the possible smoothness of the
universe (Ribeiro 2001). Bearing these points in mind, it becomes
clear that comparing $\gamma$ with $\gamma_0$ and $\gamma^\ast$
with $\gamma_0^\ast$ does not provide a true check of the
cosmological model against observations, since the assumed cosmological
model appears in {\it both} pairs of equations.\footnote{ \ This is
also true for {\it any} density constructed with galaxy samples
(Ribeiro 2001; Abdalla, Mohayaee and Ribeiro 2001).} We cannot
yet determine the geometrical factors purely observationally -- they
must be calculated from the cosmological model which we adopt on
other evidential grounds. Therefore, such a comparison is merely a
check of consistency, the best one can do at the moment considering
the present limitations in determining the characteristics and
distances of galaxies. 

\subsubsection{Cosmological Distances and the Reciprocity Law}

As stated above the observed distance $d_0$ is not unique. There are 
at least four distinct possibilities (Ellis 1971; Ribeiro 2001, 2002,
2005): the observer area distance $\da$ (which is equal to the
angular diameter distance), already introduced in \S \ref {gen} above,
the {\it luminosity distance} $\dl$, the {\it galaxy area distance}
$\dg$ (also known as {\it effective distance}, {\it angular size
distance}, {\it transverse comoving distance} or {\it proper motion
distance}) and the {\it redshift distance} $\dz$. These distances have
expressions dependent on the assumed cosmological model. In FLRW
cosmologies the last one may be written as,
\be \dz=\frac{cz}{H_0}. \lb{dz} \ee 
The other three distances are linked to each other
by the remarkable {\it reciprocity theorem} or {\it Etherington's
reciprocity law}, valid for {\it any} cosmological model and which
reads as follows (Etherington 1933; Ellis 1971; Pleba\'{n}ski and
Krasi\'{n}ski 2006, p.\ 256),
\be \dl= { \left( 1+z \right) }^2 \da = \left( 1+z \right) \dg.
    \lb{eth}
\ee

These distances can, in principle, be observationally determined,
since this law requires that source and observer should only be
connected by null geodesics (Ellis 1971). Furthermore, if we assume a
cosmological model, theoretically defined distances such as the
comoving distance, the proper distance, the distance defined by
Mattig's formula, etc, can be reduced to one of the four above
(see, {\it e.g.}, Ribeiro 2001, 2005).  

\subsection{Einstein-de Sitter Model}\lb{eds}

In EdS, besides equation (\ref{dz}) the other observational
distances are straightforwardly calculated in terms of $z$ as being
given by (Ribeiro 2001, 2005),
\be
   \dl=\frac{2c}{H_0} \left( 1+z- \sqrt{1+z} \right),
   \lb{dlz}
\ee
\be
   \da=\frac{2c}{H_0} \frac{\left( 1+z- \sqrt{1+z}
          \right)}{{ \left( 1+z \right) }^2},
   \lb{daz}
\ee
\be
   \dg= \frac{2c}{H_0} \left( \frac{ 1+z- \sqrt{1+z}}{1+z} \right),
   \lb{dgz}
\ee
and\footnote{ \ In EdS model the {\it comoving distance} is equal
to $\dg$, apart from a constant (Ribeiro 2005).} these results
allow us to derive expressions for the observed area and volume
for each distance, {\it i.e.},
$S_{\ssty L}$, $S_{\ssty A}$, $S_{\ssty G}$, $S_z$
and $V_{\ssty L}$, $V_{\ssty A}$, $V_{\ssty G}$, $V_z$
in terms of the redshift (Ribeiro 2005). Thus we are able
to use the results of table \ref{table4} to calculate the
observed differential density (\ref{gamma3}) and, after numerical
interpolations and quadratures, the observed integral differential
density (\ref{gammas3}).

Table~\ref{table5} shows the results for both differential densities
in each observational distance, as well as the values for each distance
in the redshift range of the CNOC2 sample. The figures show clearly that
the general conclusions about the dependence on the distance choice in
measuring the density of the distribution of galaxies hold. Both the
observed differential density $\gamma_0$ and the integral differential
density $\gamma_0^\ast$ {\it increase} for higher $z$ when written in
terms of the area distance $\da$, whereas both {\it decrease} when they 
are written in terms of $\dl$ or $\dz$. There is a slight increase when
$\dg$ is used, but that is mostly within the error margins and, even so,
so small for both ${ \left[ \gamma_{\ssty G} \right] }_0$ and
${ \left[ \gamma_{\ssty G}^\ast \right] }_0$ that one can consider that
in practice both density statistics {\it remain unchanged} when
written as functions of the galaxy area distance $\dg$. Since this
last distance is the same as the comoving distance (Ribeiro 2005),
widely used in testing {\it both} the possible homogeneity or
inhomogeneity of the galaxy distribution, it seems that the
conclusion reached in Ribeiro (2001) and (2005) about $\dg$ being
inappropriate for measuring the possible {\it in}homogeneity of the
Universe do seem to hold, at least as far as the CNOC2 survey is
concerned. Comoving distance is only suitable to test {\it spatial}
homogeneity in FLRW models. These results are shown graphically in
figures \ref{fig1} and \ref{fig2}. 

Figures \ref{fig3} and \ref{fig4} show both differential densities
obtained from the CNOC2 data and plotted against the luminosity and
redshift distances, respectively. $\gamma$ and $\gamma^\ast$
decrease as $z$ increases, showing a tendency towards a power law
behavior at scales of $z>0.1$. Since power laws are usually related
to self-similar patterns (Pietronero 1987; Ribeiro and Miguelote
1998; Gabrielli \etal 2005), we have drawn straight lines for reference,
whose slopes would approximately give the respective fractal dimensions
of the distribution. The plots themselves do not prove that the CNOC2
galaxy distribution is fractal for $z>0.1$, but does not entirely rule
out this possibility. In our view, this possibility clearly deserves
further investigation.

Of course, this is {\it not} saying that the {\it spatial} distribution
of galaxies may be fractal. That is fitted by an EdS, or other FLRW,
model and therefore approximately homogeneous on large scales. What it is
saying is that the {\it observationally} determined average density of
galaxies down our past light cone is possibly fractal.

\subsection{FLRW Cosmology with $\mathbf{\Omega_{m_0}=0.3}$ and
           $\mathbf{\Omega_{\Lambda_0}=0.7}$}\lb{flrw}

The differential densities can only be obtained in this cosmological
model by means of numerical integrations. Here we shall not provide
details of these calculations, which are the subject of a forthcoming
paper (Iribarrem, Ribeiro and Stoeger 2006, in preparation), but 
limit ourselves in presenting data and plots containing similar results
as in the EdS case. Suffice it to say that the methodology is precisely
the same as described in \S \ref{eds}: take the data in table \ref{table4}
and apply them to equations (\ref{gamma3}) and (\ref{gammas3}), but
evaluating numerically the geometrical part involving $d_0$, $S_0$ and
$V_0$ in this FLRW cosmology. The results for all differential densities
and distances are shown in table \ref{table6}. 

Figures \ref{fig5} and \ref{fig6} show both differential densities
against the redshift using the theoretical differential number count
$dN/dz$ obtained with the model, whereas figures \ref{fig7} and
\ref{fig8} replace $dN/dz$ with the observed ${ \left[ dN/dz \right]
}_0$ of table \ref{table4} obtained from the CNOC2 survey. 
One can clearly notice that all densities are strongly dependent on
the distance choice, not only the theoretical, but the observed ones.
However, the densities calculated with the galaxy area distance
$\dg$ change from a small decrease in the EdS model to a more
pronounced one in figures \ref{fig7} and \ref{fig8}. That decrease
does not happen in the theoretical plots shown in figures \ref{fig5}
and \ref{fig6}, meaning that it can only be attributed to the CNOC2
data. One can also notice a clear power law pattern for densities
plotted with the luminosity and redshift distances $\dl$ and $\dz$ at
ranges where the redshift is higher than $0.1$.

Finally, figures \ref{fig9} and \ref{fig10} show both differential
densities respectively plotted against $\dl$ and $\dz$. As in the EdS
model, there is a power law pattern in the tail of the plot, that is,
for $z>0.1$. That pattern seems to be even more linear than for the EdS
cosmology. Straight lines for reference have also been drawn, showing
the approximate slope of the linear decaying behavior. Since power law
patterns are usually related to self-similar fractal structures, the
possibility that some standard cosmological models possess 
{\it observational} fractal patterns at ranges of $z>0.1$ deserves
further investigation in other galaxy survey samples. 

\section{Conclusions}

In this paper we have extended the theory developed by Ribeiro and
Stoeger (2003) for linking the relativistic cosmology source count to
the luminosity function (LF) in order to extract the differential
number count $dN/dz$ from LF data. Since LF parameters assume a
cosmological model, we developed the analytical means to factor out the
assumed cosmology from the LF parameters, while preserving the LF
analytical assumptions and source evolution corrections. We have
applied this methodology to the CNOC2 galaxy redshift survey
(Lin \etal 1999) and successfully extracted the observed differential
number count ${\left[dN/dz\right]}_0$ from this galaxy catalogue.
Then we used this extracted data to carry out a statistical analysis
based on the radial differential densities discussed in Ribeiro
(2005). We have done so in the Einstein-de Sitter cosmology and the
currently favored ${\Omega_{m_0}=0.3}$, ${\Omega_{\Lambda_0}=0.7}$
standard cosmological model. Our main aim was to test Ribeiro's (2005)
proposition that different distance definitions when applied to galaxy
survey analyses introduce ambiguities that render the statistical
results unavoidably dependent on the distance choice.

Our results show that Ribeiro's (2005) proposition is sound for both
models under study. We calculated two types of radial statistical
tools, namely the differential density $\gamma$ and the integral
differential density $\gamma^\ast$, with the CNOC2 derived data and
found out that both have a strong dependency on the distance
choice, as well as, showing an {\it observational} power law pattern
for $z>0.1$ when $\gamma$ and $\gamma^\ast$ are calculated with the
luminosity distance $\dl$ and the redshift distance $\dz$.\footnote{
 \ Notice that this power law pattern only appears with {\it some}
observational quantities. As stressed in \S \ref{soh} above, the
results of this paper only refer to {\it observational} homogeneity
as the underlying models still keep their {\it spatial} homogeneity,
which is there by construction. See also Ribeiro (2001) for in depth
discussions of this topic.} This power law decay is present in both
cosmological models, EdS and ${\Omega_{m_0}=0.3}$,
${\Omega_{\Lambda_0}=0.7}$ standard model, although the linear power
law tendency is more pronounced in the latter cosmology. If we
cautiously interpret this power law pattern as possible evidence of
a self-similar fractal structure, that would mean {\it observational}
galaxy distribution fractal dimensions
in the range $D \approx 1.6-1.8$ at scales of $z > 0.1$.  

In addition to the remarks of \S \ref{soh} above, it is helpful
to remember that the cumulative number counts $N_0$ is a
dimensionless observational quantity with no dependence on any
volume definition. It can be used in other contexts, like star counting,
in a purely observational setting, having no relation to theory.
However, in determining related parameters -- $\langle n_0 \rangle$
for example -- we must know over what volume $N_0$ has been counted,
and this involves determining or assuming a definite cosmological
model. In a cosmological context the theoretical counterpart of
$N_0$, that is $N$, scales with the redshift according to the
overall mass-energy distribution of the universe, and this
theoretical dependence given by the cosmological models is what we
seek to relate to number count observations, that is, to $N_0$. 

The link between $N$ and $N_0$, or between derivative differential
number counts $dN/dz$ and ${ \left[ dN/dz \right] }_0$, is provided
by equation (\ref{um}). In this equation $dN$ is written in terms
of the proper number density $n$. Since in equation (\ref{um}) $n$
is written in terms of proper volumes, it was necessary to use the
area distance -- to tie together the locally observed cross-sectional
area, which enters in the proper volume, with the observed solid
angle. So, we can only use equation (\ref{um}) to obtain $dN/dz$
from a number density if we write $n$ in terms of the proper
density. If $n$ is not given in terms of the proper volume, a
conversion is required, that being the case when one deals with
the LF in its usual form. In theoretical calculations we can take the
local mass-energy density $\rho$ that appears on the right hand
side of Einstein's field equation and get $n$ from it (RS03, eq.\
16). Then $N=N(z)$ is independent of any volume, but still dependent
on the overall matter distribution in the universe, which, through
General Relativity, is given by the spacetime geometry. In particular,
the affine parameter $y$ is required in deriving $N$ and the
dependence of $y$ on $z$ is given by the cosmology specifically
adopted ({\it e.g.}, eq.\ \ref{dydz} above).

As a result of the procedure outlined above, the final $dN/dz$ is
dimensionless and volume independent, meaning that we are then free
to replace $dN/dz$ as given by equation (\ref{um}) with our observed,
and also dimensionless, ${ \left[ dN/dz \right] }_0$ in order to
define other number densities in different volumes. The $\gamma_0$
and $\gamma^\ast_0$ functions are just different number densities,
built in such a way as to show not only the scaling with $z$
(dependence with the overall matter distribution), but also its
dependence on one of the possible volume definitions of the assumed
cosmological model (geometrical dependence). The idea is to make
explicit the geometrical dependence on distance choice, whereas this
choice has been implicit in most discussions about the large-scale
galaxy distribution in the Universe.

We should also mention that linking the observational counts and
densities with spatial ones, and with spatial homogeneity, takes some
work. One way one can do this is by using ${ \left[ dN/dz \right] }_0$
data and Ellis' formula (\ref{um}), or its observational version
(eq.\ \ref{quatro}) (Stoeger 1987; Ara\'{u}jo and Stoeger 1999) to obtain
$n(z)$ at different redshifts, by observationally determining
$\da = \da(z)$ for at least some galaxies in the sample, and see if
this $n$ as function of $z$ agrees with the FLRW theoretical
results. This would really give us the proper densities on space-like
slices, and enable us to see {\it spatial} homogeneity. This is
quite different from integrating over large volumes down our past
light cone to get observationally averaged densities, which convolve
proper densities from a wide range of redshifts.

As a final remark, since spatial and observational homogeneities are
two different features of a cosmological model, as extensively discussed
in \S \ref{soh} above, it should be clear by now that further work
needs to be done in this area, including a clear treatment of how
one observationally tests for spatial homogeneity using observed
galaxy distributions and other data, without first demonstrating or
assuming that the cosmological region under consideration is
``almost FLRW.''

%%%%%%%%%%%%%%%%%%%%%%%%%%%%%%%%%%%%%%%%%%%%%%%%% acknowledgements
\acknowledgments
We thank the referee for very useful remarks and suggestions which
improved the paper. Two of us (V.V.L.A.\ and M.B.R.)
are grateful to CNPq and FAPERJ for financial support. 

%%%%%%%%%%%%%%%%%%%%%%%%%%%%%%%%%%%%%%%%%%%%%%%%%appendix
\appendix

\section{Error Analysis}\lb{app}

In RS03 the selection function for Schechter's type LF was written
in terms of the absolute magnitude $M_{\ssty W}$ in the filter $W$
as follows (RS03, eq.\ 96),
\be
   \psi^{\ssty W} (z) =\int^{M_{\ssty W}(z)}_{-\infty}
   \left( 0.4 \ln 10 \right) \phi_\ast 10^{0.4 \left( 1+\alpha
   \right) \left( M_\ast - {\overline{M}}_{\ssty W} \right) }
   \exp \left[ -10^{0.4 \left( M_\ast - {\overline{M}}_{\ssty W}
   \right)} \right] d \; {\overline{M}}_{\ssty W},
   \lb{psiw2}
\ee
where the three parameters that characterize the LF are
the space density of galaxies $\phi_\ast$, the asymptotic slope of the
faint end $\alpha$, and the luminosity scale given in terms of absolute
magnitude $M_\ast$. The CNOC2 survey incorporates a luminosity scale
evolution such that $M_\ast$ was rewritten in terms of two other
parameters $M_\ast^\prime$ and $Q$ (Lin \etal 1999, eq.\ 10; RS03,
eq.\ 63) leading equation (\ref{psiw2}) to be rewritten as follows,
\begin{eqnarray}
   \psi^{\ssty W} (z) & = & 0.4 \ln 10 \; \phi_\ast \lb{psiw3} \\
      & \times &  \int^{M_{\ssty W}(z)}_{-\infty} 10^{0.4 \left( 1+
      \alpha \right) \left[ M_\ast^\prime - Q \left( z-0.3 \right)
      - {\overline{M}}_{\ssty W} \right] } \exp \left\{ -10^{0.4
      \left[ M_\ast^\prime - Q \left( z-0.3 \right) -
      {\overline{M}}_{\ssty W} \right] } \right\} d \;
      {\overline{M}}_{\ssty W}. \nonumber
\end{eqnarray}
Lin \etal (1999) provided uncertainties for these four parameters,
$M_\ast^\prime$, $\alpha$, $\phi_\ast$, $Q$ in their tables 1-3. 
These errors were propagated in equation (\ref{psiw3}) by means of
standard error propagation techniques in order to produce the
uncertainties appearing in tables \ref{table1}, \ref{table2}, 
\ref{table3} of this paper. Numerical and algebraic computing
scripts were used for this task.

To obtain uncertainties for the observed differential number
counts ${[dN/dz]}_0$ shown in table \ref{table4}, we have used
equation (\ref{dndzpsi}) to further propagate the already calculated
errors for the selection functions. Similarly, by means of equation
(\ref{gamma2}) it was straightforward to evaluate uncertainties
for the differential density $\gamma$. These results are shown in
tables \ref{table5} and \ref{table6}.

To evaluate uncertainties for the integral differential density
$\gamma^\ast$ we proceeded as follows. From equation (\ref{nmed})
it is clear that $\delta \gamma^\ast = \delta N / V$. Since we can
write $N=\int (dN/dz) dz$, it is easy to find $\delta N$, and,
therefore, uncertainties of the integral differential density yield,
\be
   \delta \gamma^\ast= \left( \frac{1}{V} \right) { \left( \frac{d^{\:
                        2}N}{dz^2} \right) }^{-1} \left( \frac{dN}{dz}
                       \right) \; \delta \left( dN/dz \right).
   \lb{deltagammas}
\ee
Equation (\ref{dndzpsi}) and the already tabulated errors of the observed
differential number counts $\delta [ dN/dz ]_0$ allowed us to obtain the
values for $\delta \gamma^\ast_0$. The results are shown in tables
\ref{table5} and \ref{table6}.

%%%%%%%%%%%%%%%%%%%%%%%%%%%%%%%%%%%%%%%%%%%%%%%%% References
%% cross-referencing. Each reference is preceded by a \bibitem command
%% that defines in curly braces the KEY that corresponds to the KEY in
%% the \cite commands. Make sure that you provide a unique KEY for
%% every \bibitem or else the paper will not LaTeX. The square brackets
%% should contain the citation text that LaTeX will insert in place of
%% the \cite commands.

\clearpage
% table 1
\begin{deluxetable}{cccc}
\tabletypesize{\normalsize}
\tablecaption{CNOC2 selection function\tablenotemark{a} \ results
              from RS03 plus errors evaluated here (see appendix
	      \ref{app}) in the $R_c$ band  vs.\ redshift.\lb{table1}}
\tablewidth{0pt}
\tablehead{
\colhead{$z$} & \colhead{$\psi_1^{\ssty {R_c}}$} &
\colhead{$\psi_2^{\ssty {R_c}}$} & \colhead{$\psi_3^{\ssty {R_c}}$}
}
\startdata
  0.05 & 0.0181 $\pm$ 0.0038& 0.0120 $\pm$ 0.0040& 0.0220 $\pm$ 0.0130\\ 
  0.12 & 0.0182 $\pm$ 0.0038& 0.0121 $\pm$ 0.0041& 0.0221 $\pm$ 0.0130\\ 
  0.25 & 0.0183 $\pm$ 0.0039& 0.0123 $\pm$ 0.0041& 0.0223 $\pm$ 0.0130\\ 
  0.4  & 0.0185 $\pm$ 0.0039& 0.0125 $\pm$ 0.0042& 0.0225 $\pm$ 0.0132\\ 
  0.55 & 0.0186 $\pm$ 0.0040& 0.0127 $\pm$ 0.0044& 0.0227 $\pm$ 0.0135\\ 
  0.75 & 0.0187 $\pm$ 0.0040& 0.0130 $\pm$ 0.0045& 0.0230 $\pm$ 0.0141\\ 
  0.9  & 0.0187 $\pm$ 0.0041& 0.0132 $\pm$ 0.0046& 0.0232 $\pm$ 0.0148\\ 
  1.0  & 0.0187 $\pm$ 0.0041& 0.0133 $\pm$ 0.0047& 0.0234 $\pm$ 0.0153\\ 
\enddata
\tablenotetext{a}{Subscript numbers denote the spectral type
                  morphology adopted for the CNOC2 survey
		  (see \S \ref{cnoc2} above). Units are
		  Mpc$^{-3}$.} % h$^3$.}
\end{deluxetable}
\clearpage 
% table 2
\begin{deluxetable}{cccc}
\tabletypesize{\normalsize}
\tablecaption{CNOC2 selection function results from RS03 plus errors
              evaluated here (see appendix \ref{app}) in the
              $U$ band vs.\ redshift.\lb{table2}}
\tablewidth{0pt}
\tablehead{
\colhead{$z$} & \colhead{$\psi_1^{\ssty U}$} &
\colhead{$\psi_2^{\ssty U}$} & \colhead{$\psi_3^{\ssty U}$} 
}
\startdata
  0.05 & 0.0180 $\pm$ 0.0031 & 0.0119 $\pm$ 0.0038 & 0.0284 $\pm$ 0.0134 \\ 
  0.12 & 0.0182 $\pm$ 0.0031 & 0.0121 $\pm$ 0.0038 & 0.0289 $\pm$ 0.0135 \\ 
  0.25 & 0.0186 $\pm$ 0.0032 & 0.0123 $\pm$ 0.0039 & 0.0297 $\pm$ 0.0139 \\ 
  0.4  & 0.0189 $\pm$ 0.0033 & 0.0126 $\pm$ 0.0040 & 0.0307 $\pm$ 0.0145 \\ 
  0.55 & 0.0192 $\pm$ 0.0033 & 0.0128 $\pm$ 0.0042 & 0.0317 $\pm$ 0.0152 \\ 
  0.75 & 0.0194 $\pm$ 0.0034 & 0.0131 $\pm$ 0.0043 & 0.0330 $\pm$ 0.0162 \\ 
  0.9  & 0.0195 $\pm$ 0.0034 & 0.0134 $\pm$ 0.0045 & 0.0341 $\pm$ 0.0172 \\ 
  1.0  & 0.0196 $\pm$ 0.0034 & 0.0135 $\pm$ 0.0045 & 0.0348 $\pm$ 0.0179 \\ 
\enddata
\end{deluxetable}
\clearpage 
% table 3
\begin{deluxetable}{cccc}
\tabletypesize{\normalsize}
\tablecaption{CNOC2 selection function results from RS03 plus errors
              evaluated here (see appendix \ref{app}) in the
              $B_{\ssty AB}$ band vs.\ redshift.\lb{table3}}
\tablewidth{0pt}
\tablehead{
\colhead{$z$} & \colhead{$\psi_1^{\ssty {B}}$} &
\colhead{$\psi_2^{\ssty {B}}$} & \colhead{$\psi_3^{\ssty {B}}$}
}
\startdata
  0.05 & 0.0182 $\pm$ 0.0033 & 0.0122 $\pm$ 0.0036 & 0.0252 $\pm$ 0.0130 \\ 
  0.12 & 0.0183 $\pm$ 0.0033 & 0.0123 $\pm$ 0.0036 & 0.0254 $\pm$ 0.0130 \\ 
  0.25 & 0.0185 $\pm$ 0.0034 & 0.0125 $\pm$ 0.0037 & 0.0257 $\pm$ 0.0131 \\ 
  0.4  & 0.0187 $\pm$ 0.0034 & 0.0128 $\pm$ 0.0038 & 0.0260 $\pm$ 0.0133 \\ 
  0.55 & 0.0189 $\pm$ 0.0035 & 0.0130 $\pm$ 0.0040 & 0.0264 $\pm$ 0.0136 \\ 
  0.75 & 0.0190 $\pm$ 0.0035 & 0.0133 $\pm$ 0.0041 & 0.0268 $\pm$ 0.0143 \\ 
  0.9  & 0.0191 $\pm$ 0.0035 & 0.0135 $\pm$ 0.0042 & 0.0272 $\pm$ 0.0150 \\ 
  1.0  & 0.0191 $\pm$ 0.0035 & 0.0136 $\pm$ 0.0043 & 0.0274 $\pm$ 0.0155 \\ 
\enddata
\end{deluxetable}
\clearpage 
% table 4
\begin{deluxetable}{cc}
\tabletypesize{\normalsize}
\tablecaption{CNOC2 differential counts for all wave bands and morphologies
              vs.\ redshift.\lb{table4}}
\tablewidth{0pt}
\tablehead{
\colhead{$z$} & \colhead{$ { \left[
		\frac{\displaystyle dN}{\displaystyle dz} \right] }_0$} 
\vspace{2.5mm}}
\startdata
  0.05 & (2.914 $\pm$ 1.024) $\times 10^{7}$  \\ 
  0.12 & (1.394 $\pm$ 0.491) $\times 10^{8}$  \\ 
  0.25 & (4.409 $\pm$ 1.564) $\times 10^{8}$  \\ 
  0.4  & (8.125 $\pm$ 2.920) $\times 10^{8}$  \\ 
  0.55 & (1.142 $\pm$ 0.417) $\times 10^{9}$  \\ 
  0.75 & (1.487 $\pm$ 0.555) $\times 10^{9}$  \\ 
  0.9  & (1.679 $\pm$ 0.639) $\times 10^{9}$  \\ 
  1.0  & (1.780 $\pm$ 0.686) $\times 10^{9}$  \\ 
\enddata
\end{deluxetable}
\clearpage
% table 5
\thispagestyle{empty}
\begin{deluxetable}{lrrrrrrrrrrrr}
\rotate
\tabletypesize{\scriptsize}
\tablecaption{Redshifts, distances and differential densities calculated
              with CNOC2 differential number count data in an Einstein-de Sitter
              cosmology. Distances are in Mpc, densities in $\mathrm{Mpc}^{-3}$
              and $H_0=100 \; \mathrm{km} \; \mathrm{s}^{-1} \;
              \mathrm{Mpc}^{-1}$. The differential densities were
	      calculated using equations (\protect\ref{gamma3}) and
	      (\protect\ref{gammas3}), where the observed differential
	      number count ${ \left[ dN/dz \right] }_0$
	      was given by the survey data results shown in table
	      \protect\ref{table4}, whereas the second factor in the
	      right hand side of equation (\protect\ref{gamma3}) was
	      obtained theoretically, that is, from the assumed
	      cosmological model.\lb{table5}}
\tablewidth{0pt}
\setlength{\tabcolsep}{1.8mm}
\tablehead{
\colhead{$z$} &
\colhead{$\da$} &
\colhead{$\dg$} &
\colhead{$\dl$} &
\colhead{$\dz$} &
\colhead{${ \left[ \gamma_{\ssty A} \right] }_0$} &
\colhead{${ \left[ \gamma_{\ssty G} \right] }_0$} &
\colhead{${ \left[ \gamma_{\ssty L} \right] }_0$} &
\colhead{${ \left[ \gamma_z \right] }_0$}&
\colhead{${ \left[ \gamma_{\ssty A}^\ast \right] }_0$} &
\colhead{${ \left[ \gamma_{\ssty G}^\ast \right] }_0$} &
\colhead{${ \left[ \gamma_{\ssty L}^\ast \right] }_0$} &
\colhead{${ \left[ \gamma_z^\ast \right] }_0$} 
}
\startdata
0.05 & 138 &  144 &  152 &  150 & 0.0485$\pm$0.0050 & 0.0399$\pm$0.0041 & 0.0328$\pm$0.0034 & 0.0344$\pm$0.0036& 0.0437$\pm$0.0074 & 0.0377$\pm$0.0064 & 0.0326$\pm$0.0055 & 0.0338$\pm$0.0057 \\ 
0.12 & 295 &  330 &  370 &  360 & 0.0640$\pm$0.0066 & 0.0402$\pm$0.0041 & 0.0256$\pm$0.0026 & 0.0286$\pm$0.0029& 0.0561$\pm$0.0095 & 0.0399$\pm$0.0068 & 0.0284$\pm$0.0048 & 0.0309$\pm$0.0052 \\ 
0.25 & 506 &  633 &  791 &  749 & 0.1044$\pm$0.0107 & 0.0408$\pm$0.0042 & 0.0169$\pm$0.0017 & 0.0208$\pm$0.0021& 0.0791$\pm$0.0147 & 0.0405$\pm$0.0075 & 0.0207$\pm$0.0038 & 0.0244$\pm$0.0045 \\ 
0.4  & 663 &  928 & 1300 & 1199 & 0.1795$\pm$0.0186 & 0.0414$\pm$0.0043 & 0.0111$\pm$0.0011 & 0.0150$\pm$0.0016& 0.1124$\pm$0.0236 & 0.0410$\pm$0.0086 & 0.0149$\pm$0.0031 & 0.0190$\pm$0.0040 \\ 
0.55 & 761 & 1180 & 1829 & 1649 & 0.3067$\pm$0.0322 & 0.0420$\pm$0.0044 & 0.0076$\pm$0.0008 & 0.0111$\pm$0.0012& 0.1540$\pm$0.0370 & 0.0414$\pm$0.0099 & 0.0111$\pm$0.0027 & 0.0152$\pm$0.0036 \\ 
0.75 & 836 & 1463 & 2561 & 2248 & 0.6456$\pm$0.0694 & 0.0427$\pm$0.0046 & 0.0048$\pm$0.0005 & 0.0078$\pm$0.0008& 0.2242$\pm$0.0663 & 0.0418$\pm$0.0124 & 0.0078$\pm$0.0023 & 0.0115$\pm$0.0034 \\ 
0.9  & 866 & 1646 & 3127 & 2698 & 1.2154$\pm$0.1339 & 0.0431$\pm$0.0047 & 0.0036$\pm$0.0004 & 0.0061$\pm$0.0007& 0.2891$\pm$0.1029 & 0.0422$\pm$0.0150 & 0.0061$\pm$0.0022 & 0.0096$\pm$0.0034 \\ 
1.0  & 878 & 1756 & 3512 & 2998 & 2.0200$\pm$0.2268 & 0.0433$\pm$0.0049 & 0.0030$\pm$0.0003 & 0.0053$\pm$0.0006& 0.3387$\pm$0.1398 & 0.0423$\pm$0.0175 & 0.0053$\pm$0.0022 & 0.0085$\pm$0.0035 \\ 
\enddata
\end{deluxetable}
\clearpage
% table 6
\thispagestyle{empty}
\begin{deluxetable}{lrrrrrrrrrrrr}
\rotate
\tabletypesize{\scriptsize}
\tablecaption{Redshifts, distances and differential densities calculated
              with CNOC2 differential number count data in a
	      $\Omega_{m_0}=0.3$, $\Omega_{\Lambda_0}=0.7$ standard
              cosmology. Distances are in Mpc, densities in $\mathrm{Mpc}^{-3}$
              and $H_0=100 \; \mathrm{km} \; \mathrm{s}^{-1} \;
              \mathrm{Mpc}^{-1}$. The differential densities were
	      calculated similarly as explained in the caption of
	      table \protect\ref{table5}.\lb{table6}}
\tablewidth{0pt}
\setlength{\tabcolsep}{1.5mm}
\tablehead{
\colhead{$z$} &
\colhead{$\da$} &
\colhead{$\dg$} &
\colhead{$\dl$} &
\colhead{$\dz$} &
\colhead{${ \left[ \gamma_{\ssty A} \right] }_0$} &
\colhead{${ \left[ \gamma_{\ssty G} \right] }_0$} &
\colhead{${ \left[ \gamma_{\ssty L} \right] }_0$} &
\colhead{${ \left[ \gamma_z \right] }_0$}&
\colhead{${ \left[ \gamma_{\ssty A}^\ast \right] }_0$} &
\colhead{${ \left[ \gamma_{\ssty G}^\ast \right] }_0$} &
\colhead{${ \left[ \gamma_{\ssty L}^\ast \right] }_0$} &
\colhead{${ \left[ \gamma_z^\ast \right] }_0$} 
}
\startdata
0.05&141 &148 &156 &150 &$0.0438 \pm 0.0045$ &$0.0360 \pm 0.0037$ &$0.0297 \pm 0.0031$ &$0.0344 \pm 0.0036$&$0.0405 \pm 0.0069$ &$0.0350 \pm 0.0059$ &$0.0302 \pm 0.0051$ &$0.0338 \pm 0.0057$ \\ 
0.12&312 &350 &392 &360 &$0.0506 \pm 0.0052$ &$0.0320 \pm 0.0033$ &$0.0205 \pm 0.0021$ &$0.0286 \pm 0.0029$&$0.0472 \pm 0.0080$ &$0.0336 \pm 0.0057$ &$0.0239 \pm 0.0040$ &$0.0309 \pm 0.0052$\\ 
0.25&565 &706 &882 &749 &$0.0662 \pm 0.0068$ &$0.0266 \pm 0.0027$ &$0.0112 \pm 0.0012$ &$0.0208 \pm 0.0021$&$0.0571 \pm 0.0106$ &$0.0292 \pm 0.0054$ &$0.0150 \pm 0.0028$ &$0.0244 \pm 0.0045$\\ 
0.4 &776 &1086&1521&1199&$0.0910 \pm 0.0094$ &$0.0226 \pm 0.0023$ &$0.0062 \pm 0.0006$ &$0.0150 \pm 0.0016$&$0.0702 \pm 0.0147$ &$0.0256 \pm 0.0054$ &$0.0093 \pm 0.0020$ &$0.0190 \pm 0.0040$\\ 
0.55&926 &1435&2224&1649&$0.1266 \pm 0.0133$ &$0.0198 \pm 0.0021$ &$0.0038 \pm 0.0004$ &$0.0111 \pm 0.0012$&$0.0856 \pm 0.0206$ &$0.0230 \pm 0.0055$ &$0.0062 \pm 0.0015$ &$0.0152 \pm 0.0036$\\ 
0.75&1060&1854&3245&2248&$0.2019 \pm 0.0217$ &$0.0174 \pm 0.0019$ &$0.0021 \pm 0.0002$ &$0.0078 \pm 0.0008$&$0.1102 \pm 0.0326$ &$0.0206 \pm 0.0061$ &$0.0038 \pm 0.0011$ &$0.0115 \pm 0.0034$\\ 
0.9 &1125&2137&4061&2698&$0.2949 \pm 0.0325$ &$0.0162 \pm 0.0018$ &$0.0015 \pm 0.0002$ &$0.0061 \pm 0.0007$&$0.1320 \pm 0.0470$ &$0.0193 \pm 0.0069$ &$0.0028 \pm 0.0010$ &$0.0096 \pm 0.0034$\\ 
1.0 &1156&2313&4625&2998&$0.3877 \pm 0.0435$ &$0.0156 \pm 0.0017$ &$0.0012 \pm 0.0001$ &$0.0053 \pm 0.0006$&$0.1483 \pm 0.0612$ &$0.0185 \pm 0.0077$ &$0.0023 \pm 0.0010$ &$0.0085 \pm 0.0035$\\ 
\enddata
\end{deluxetable}
\clearpage
\begin{figure}
\centering
%\input{fig1.tex}
%\plotone{f1.ps}
%\includegraphics[width=12cm,height=17cm,angle=-90]{f1.ps}
\includegraphics[scale=0.65,angle=-90]{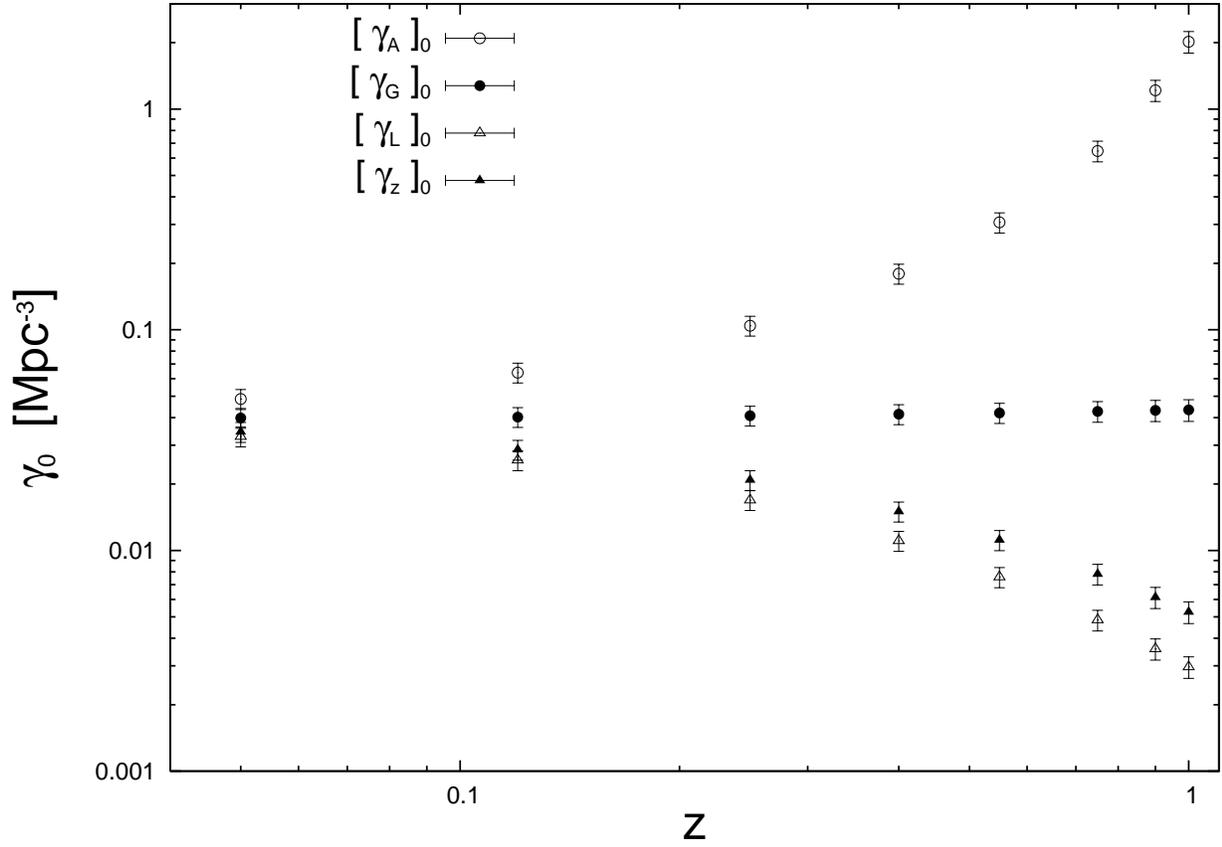}
\vspace{1cm}
\caption{Differential densities $\gamma_0$ against redshift in the EdS
         model with CNOC2 data. The same theoretical behavior of the
	 differential count presented in Ribeiro (2005) is preserved
	 with the CNOC2 differential number counting data. Notice the
	 power law pattern when $z>0.1$ for densities obtained with
	 $\dl$ and $\dz$.}\lb{fig1}
\end{figure}
\clearpage
\begin{figure}
\centering
%\input{fig2.tex}
%\plotone{f2.ps}
\includegraphics[scale=0.65,angle=-90]{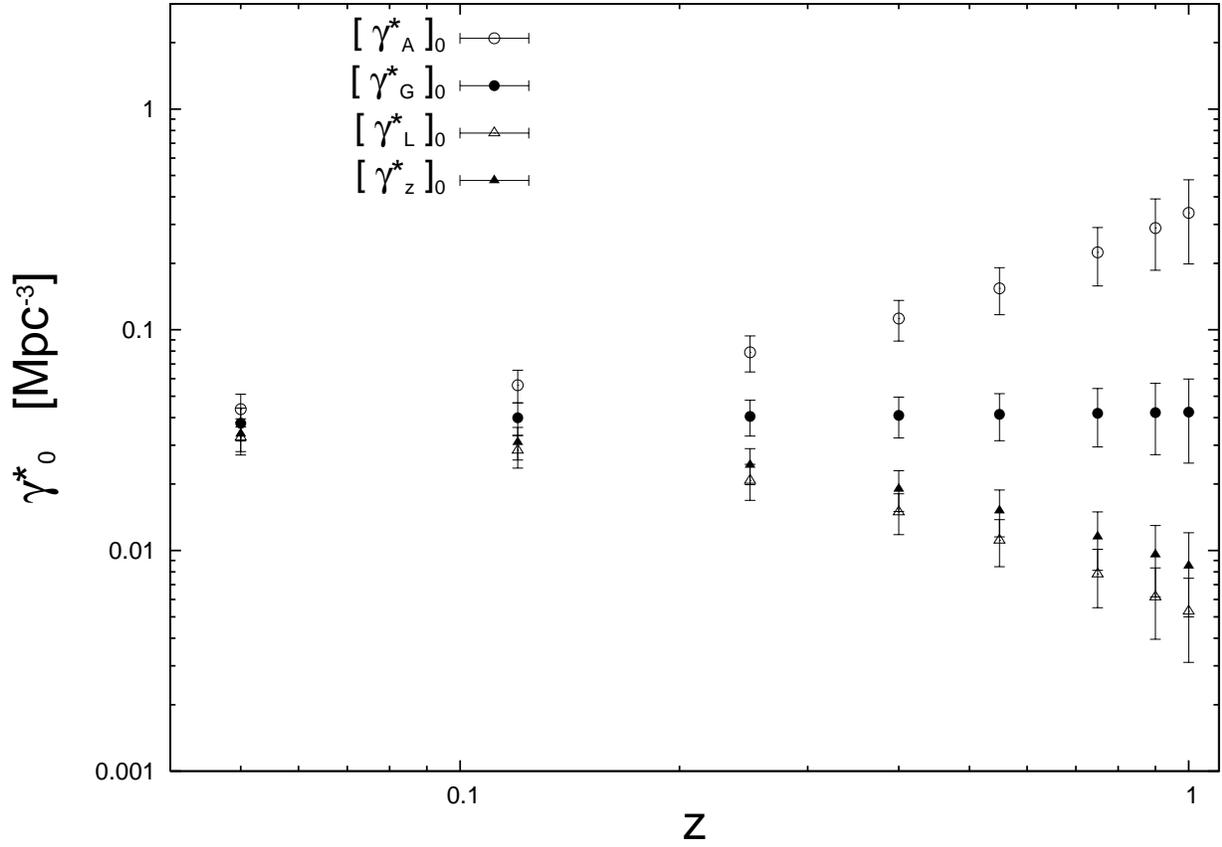}
\vspace{1cm}
\caption{Integral differential densities $\gamma_0^\ast$ against redshift
         in the EdS model. As in the previous plot, the critical dependence
	 on the choice of distance is preserved in the CNOC2
	 data, as well as the visible power law decaying pattern
	 occurring when one chooses $\dl$ and $\dz$ as distances.}\lb{fig2}
\end{figure}
\clearpage
\begin{figure}
\centering
%\input{fig3.tex}
%\plotone{f3.ps}
\includegraphics[scale=0.65,angle=-90]{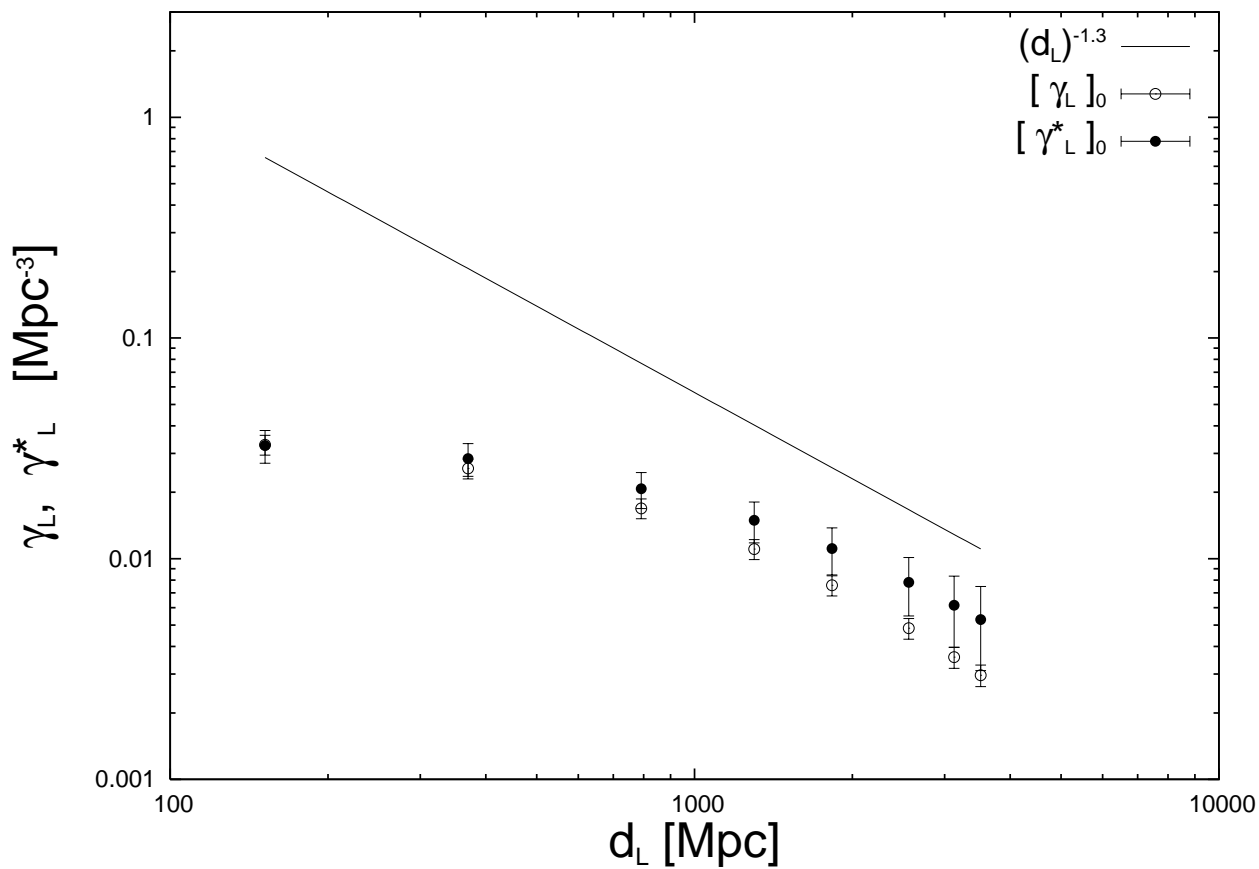}
\vspace{1cm}
\caption{Plot of the differential densities obtained from the CNOC2
        survey against the luminosity
        distance in EdS cosmology. Both densities decrease at higher
	$\dl$ and show a clear power law behavior in the
	tail. The line is for reference only, but its slope is such
	that in the language of fractals it would mean a fractal
	dimension $D=1.7$. This plot does {\it not} prove that the
	galaxy distribution as presented in the CNOC2 survey data
	follows a fractal pattern, but that such an {\it
	observationally} self-similar structure is possible at
	least in some scales.}\lb{fig3}
\end{figure}
\clearpage
\begin{figure}
\centering
%\input{fig4.tex}
%\plotone{f4.ps}
\includegraphics[scale=0.65,angle=-90]{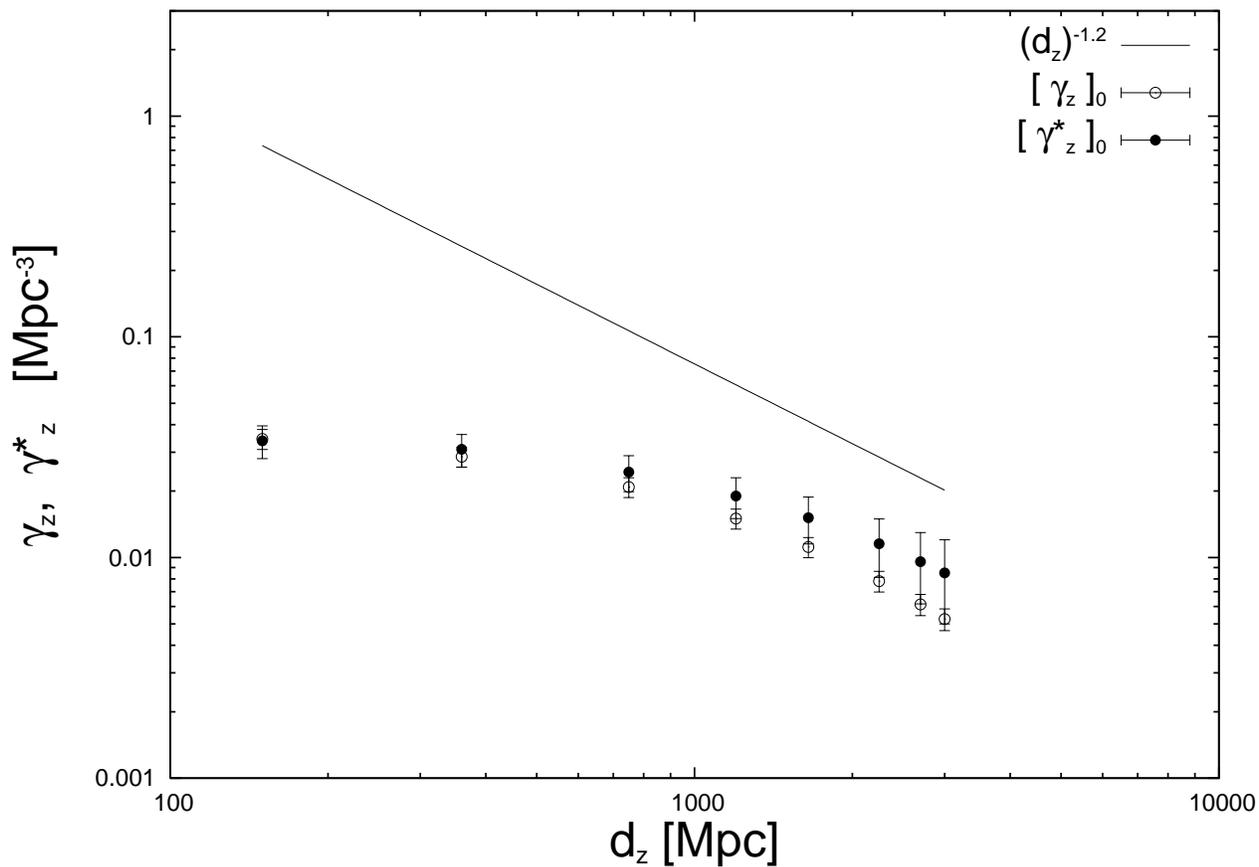}
\vspace{1cm}
\caption{Plot of the differential densities obtained from the CNOC2
         survey against the redshift
         distance in EdS cosmology. Similarly to the previous graph,
	 both densities tend to follow a power law decay at higher
	 distances. The straight line is for reference only, but its
	 slope is such that it would mean $D=1.8$. Again, one must
	 remain cautious regarding interpreting this self-similar
	 behavior in the tail as evidence of a fractal pattern.
	 However, the plot indicates that this possibility should
	 not be ruled out either. Further investigation with other
	 samples is clearly required to clarify this point.}\lb{fig4}
\end{figure}
\clearpage
\begin{figure}
\centering
%\input{fig5.tex}
%\plotone{f5.ps}
\includegraphics[scale=0.65,angle=-90]{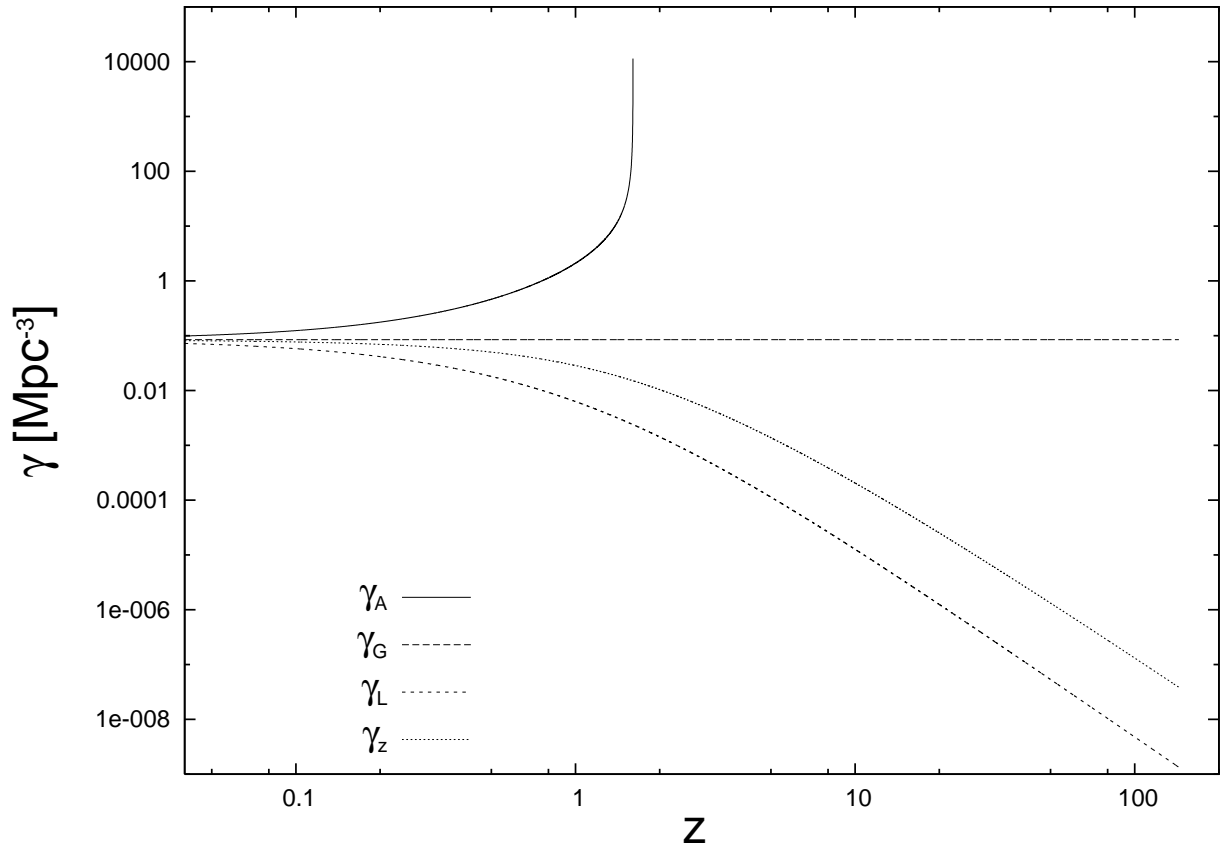}
\vspace{1cm}
\caption{Plot of the theoretical differential density $\gamma$ for each
         observational distance in the ${\Omega_{m_0}=0.3}$,
	 ${\Omega_{\Lambda_0}=0.7}$ standard cosmology. The results are
	 similar to those of EdS model as shown in Ribeiro (2005) and are
	 presented here for comparison with the observationally
	 derived $\gamma_0$ shown in figure \protect\ref{fig7}. Notice
	 that similarly to the EdS case $\gamma_{\ssty A}$ blows up at
	 about $z \approx 1.5$, but that in itself does not mean that
	 the volume defined with $\da$ is incorrect, but only that the
	 way $\gamma_{\ssty A}$ is defined has a singularity, which has
	 no special physical significance other than that the distance
	 it involves has a critical point.}\lb{fig5}
\end{figure}
\clearpage
\begin{figure}
\centering
%\input{fig6.tex}
%\plotone{f6.ps}
\includegraphics[scale=0.65,angle=-90]{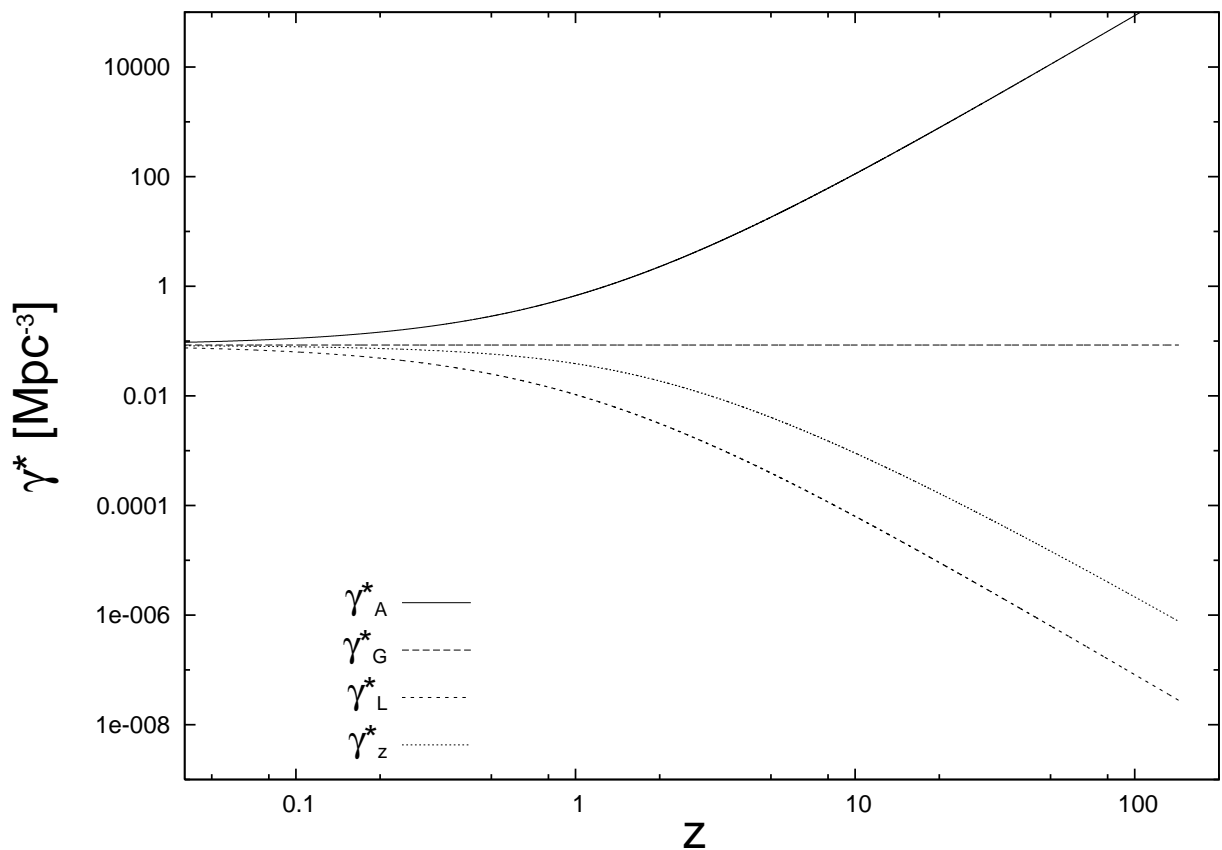}
\vspace{1cm}
\caption{Plot of the theoretical integral differential density
         $\gamma^\ast$ for each observational distance in the
	 ${\Omega_{m_0}=0.3}$, ${\Omega_{\Lambda_0}=0.7}$ standard
	 cosmological model. The results are similar to those of EdS
	 cosmology (Ribeiro 2005). They are presented here so
	 that one can compare with the observationally derived
	 $\gamma_0^\ast$ plotted in figure \protect\ref{fig8}.}\lb{fig6}
\end{figure}
\clearpage
\begin{figure}
\centering
%\input{fig7.tex}
%\plotone{f7.ps}
\includegraphics[scale=0.65,angle=-90]{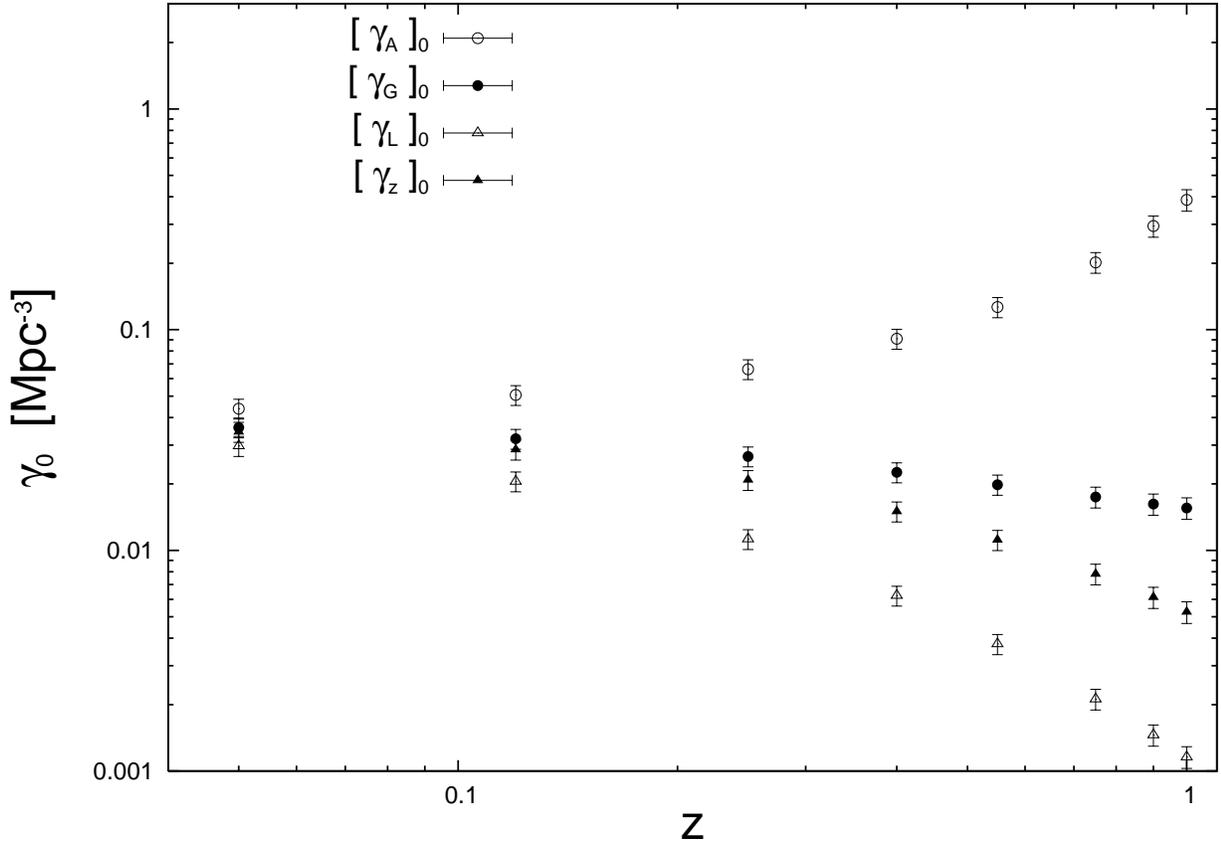}
\vspace{1cm}
\caption{This figure shows the observed differential density $\gamma_0$
         vs.\ the redshift in the ${\Omega_{m_0}=0.3}$,
	 ${\Omega_{\Lambda_0}=0.7}$ standard cosmology with CNOC2
	 differential number counts data of table
	 \protect\ref{table4}. Notice that similarly to the results
	 shown in figure \protect\ref{fig1} there is as power law
	 pattern at the tail of the plot for both $\dl$ and $\dz$.
	 However, the differential density calculated with the galaxy
	 area distance $\dg$ does not remain unchanged, but also decays
	 at higher $z$, an effect that can only be explained because
	 this plot uses the CNOC2 differential number count ${ \left[
	 dN/dz \right] }_0$ for calculating $\gamma_0$, rather than its
	 theoretically derived values used in figure~\protect\ref{fig5}.
	 Therefore, it is possible that either the CNOC2 survey is
	 undercounting galaxies at larger $z$ or that its LF source
	 evolution equation is not correct. A third, less likely,
	 possibility is that there could be real spatial inhomogeneity
	 and in this case the FLRW model would not be correct.}\lb{fig7}
\end{figure}
\clearpage
\begin{figure}
\centering
%\input{fig8.tex}
%\plotone{f8.ps}
\includegraphics[scale=0.65,angle=-90]{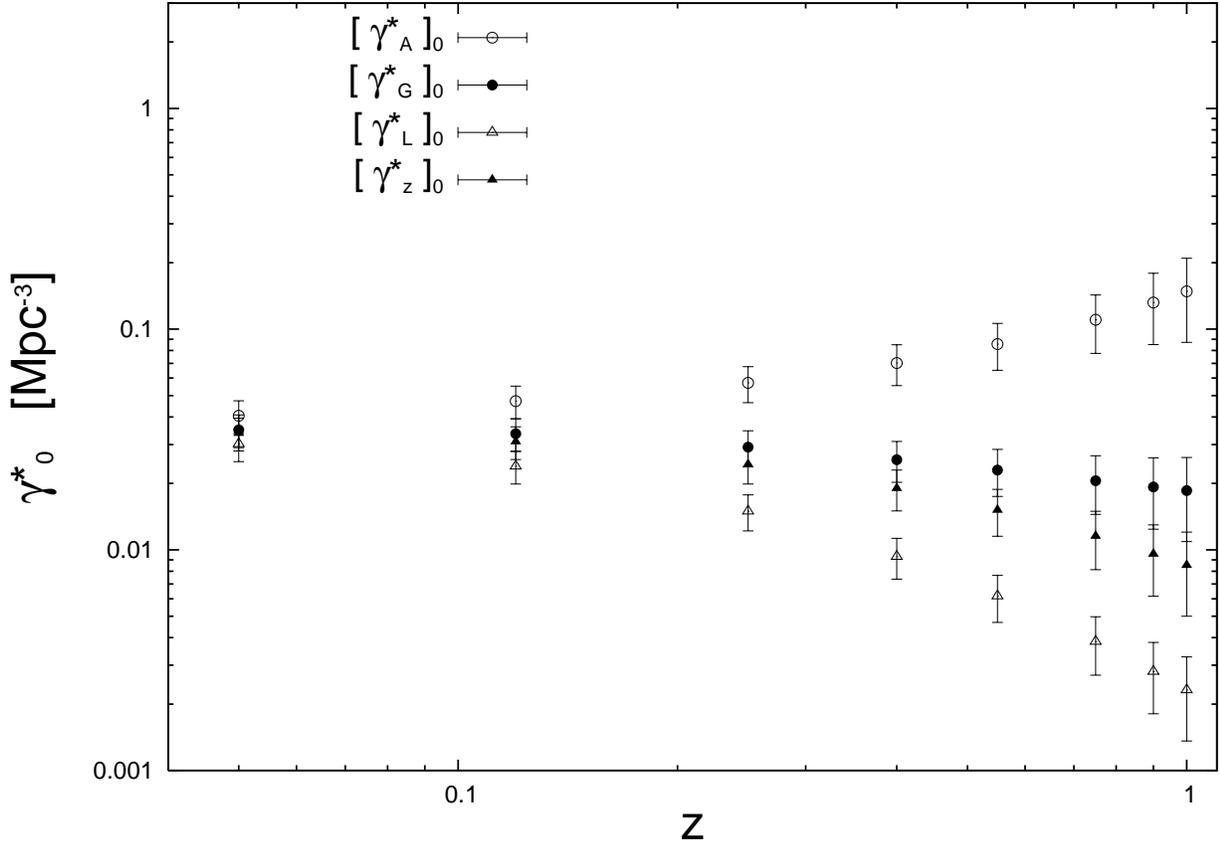}
\vspace{1cm}
\caption{Plot of the observed integral differential density
         $\gamma_0^\ast$ against the redshift in the ${\Omega_{m_0}=0.3}$,
	 ${\Omega_{\Lambda_0}=0.7}$ standard cosmology with a CNOC2 LF
	 derived ${ \left[ dN/dz \right] }_0$. As in the previous plot
	 one can also notice a power law pattern for both the luminosity
	 distance $\dl$ and redshift distance $\dz$. But, still as in
	 the previous plot, $\gamma_0^\ast$ calculated with $\dg$
	 decreases at higher $z$, departing from the theoretical
	 behavior presented in figure \protect\ref{fig6}.}\lb{fig8}
\end{figure}
\clearpage
\begin{figure}
\centering
%\input{fig9.tex}
%\plotone{f9.ps}
\includegraphics[scale=0.65,angle=-90]{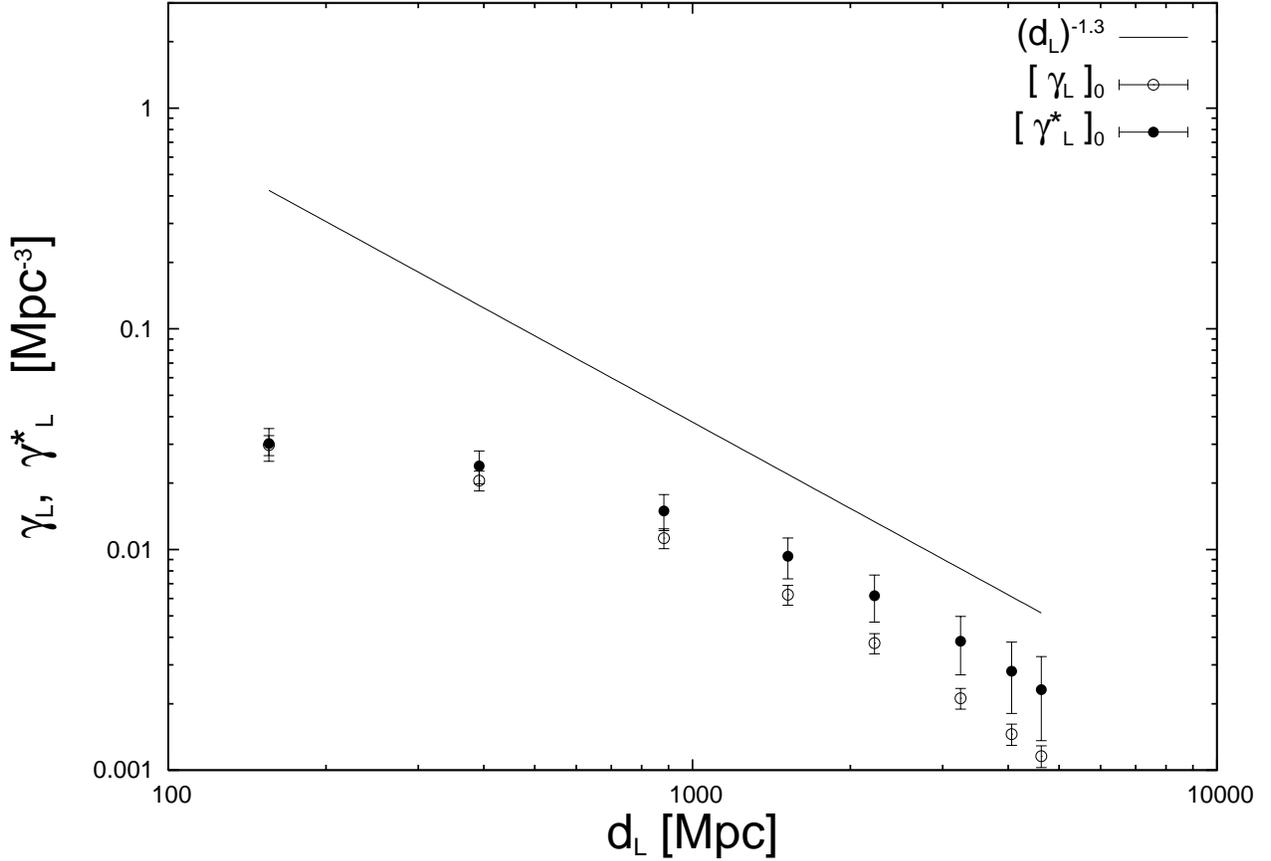}
\vspace{1cm}
\caption{Differential densities against the luminosity distance in the
         ${\Omega_{m_0}=0.3}$, ${\Omega_{\Lambda_0}=0.7}$ standard
	 cosmology using the CNOC2 differential number count data of
	 table \protect\ref{table4}. The power law behavior is
	 visible in the $z>0.1$ ranges. The straight line is for reference
	 only, but its slope would mean a fractal dimension of
	 approximately $D=1.7$. As cautioned above, one must remain
	 careful in interpreting this pattern as evidence of a fractal
	 structure, since the results were obtained only with the
	 CNOC2 survey data.}\lb{fig9}
\end{figure}
\clearpage
\begin{figure}
\centering
%\input{fig10.tex}
%\plotone{f10.ps}
\includegraphics[scale=0.65,angle=-90]{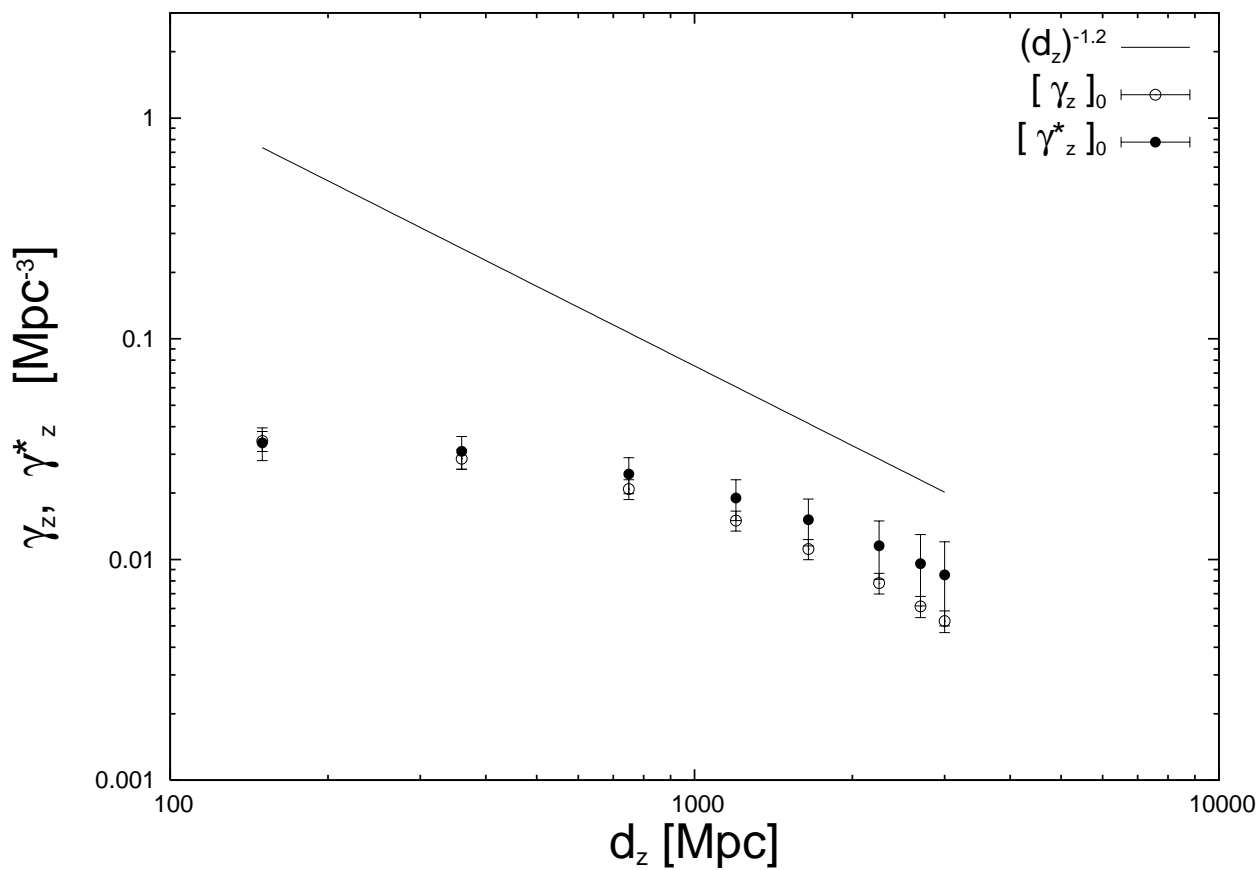}
\vspace{1cm}
\caption{Differential densities versus the redshift in the
         ${\Omega_{m_0}=0.3}$, ${\Omega_{\Lambda_0}=0.7}$ standard
	 cosmological model where ${ \left[ dN/dz \right] }_0$ was
	 derived from the LF obtained with the CNOC2 galaxy survey
	 (see table \protect\ref{table4}). Similarly to the previous
	 figure, one can notice a power law pattern in the tail of
	 both plots. The straight line is for reference only, but
	 would mean approximately $D=1.8$ if one interprets these
	 results as possible evidence of a fractal structure for $z>0.1$
	 ranges.}\lb{fig10}
\end{figure}
\end{document}